\documentclass[a4paper,11pt]{article}

\usepackage{jheppub,bm} 

\usepackage[parfill]{parskip}
\usepackage{siunitx}
\usepackage{cleveref}
\usepackage{todonotes}
\usepackage{scalefnt}
\usepackage{slashed}
\usepackage{microtype}
\usepackage{multirow}
\usepackage{csquotes}
\usepackage{verbatim}

\usepackage{caption}
\usepackage{subcaption}
\usepackage{graphicx}
\usepackage{nicefrac}

\newcommand{\abbrev}{\scalefont{.9}}

\newcommand{\NNLO}{\text{\abbrev NNLO}}
\newcommand{\NLO}{\text{\abbrev NLO}}
\newcommand{\LO}{\text{\abbrev LO}}

\newcommand{\SM}{\text{\abbrev SM}}
\newcommand{\BSM}{\text{\abbrev BSM}}

\newcommand{\QCD}{\text{\abbrev QCD}}
\newcommand{\EW}{\text{\abbrev EW}}
\newcommand{\PDF}{\text{\abbrev PDF}}
\newcommand{\MMHT}{\text{\abbrev MMHT}}
\newcommand{\LHC}{\text{\abbrev LHC}}

\newcommand{\MCFM}{\text{\abbrev MCFM}}
\newcommand{\MCFMEIGHT}{\text{\abbrev MCFM-8.0}}
\newcommand{\OL}{\text{\abbrev OpenLoops}}
\newcommand{\MG}{\text{\abbrev MadGraph}}
\newcommand{\ATLAS}{\text{\abbrev ATLAS}}
\newcommand{\CMS}{\text{\abbrev CMS}}
\newcommand{\CDF}{\text{\abbrev CDF}}
\newcommand{\DO}{\text{\abbrev D0}}
\newcommand{\SCET}{\text{\abbrev SCET}}

\newcommand{\abs}[1]{\lvert#1\rvert}

\newcommand{\taucut}{\tau_\text{cut}}
\newcommand{\pt}{p_{\text{T}}}

\allowdisplaybreaks[4]

\begin{document}

\title{$Z\gamma$ production at NNLO including anomalous couplings}

\author[]{John M. Campbell$^a$, }
\author[]{Tobias Neumann$^b$ }
\author[]{and Ciaran Williams$^b$}

\affiliation[a]{Fermilab,\\PO Box 500, Batavia, IL 60510, USA}
\affiliation[b]{Department of Physics,\\ University at Buffalo, The State University of New York, Buffalo
14260, USA}

\emailAdd{johnmc@fnal.gov}
\emailAdd{tobiasne@buffalo.edu}
\emailAdd{ciaranwi@buffalo.edu}

\abstract{
In this paper we present a next-to-next-to-leading order (\NNLO{}) \QCD{} calculation of the processes 
$pp\rightarrow l^+l^-\gamma$ and $pp\rightarrow \nu\bar\nu\gamma$ that we have implemented in 
 \MCFM{}. Our calculation includes \QCD{} corrections at \NNLO{} both for the Standard Model (\SM{}) and additionally 
 in the presence of  $Z\gamma\gamma$ and $ZZ\gamma$ 
anomalous couplings.
We compare our implementation, obtained using the jettiness
slicing approach, with a previous \SM{} calculation
and find broad agreement.
Focusing on the sensitivity of our results to the slicing parameter, 
we show that using our setup we are able to compute \NNLO{} cross sections with numerical 
uncertainties of about $0.1\%$, which is small compared to residual scale uncertainties of 
a few percent. We study potential improvements using two different jettiness definitions and the 
inclusion of power corrections. At $\sqrt{s}=\SI{13}{TeV}$ we present 
phenomenological results and consider $Z\gamma$ as a background to $H\to Z\gamma$ production.
We find that, with typical cuts, the inclusion of \NNLO{} corrections represents a small effect 
and loosens the extraction of limits on anomalous couplings by about $10\%$.
}

\preprint{FERMILAB-PUB-17-303-T}
\maketitle
\flushbottom

\section{Introduction}

The power of the Large Hadron Collider (\LHC{}) is its ability both to search for new phenomena at 
the highest energies and to accumulate a wealth of precise data on a broad range of final 
states at lower energies. The former is enabled by the center of mass energy of \SI{13}{TeV} while 
the latter relies on both the immense amounts of data collected and the technical capabilities of 
the detector, as well as the ingenuity of the experimental analyses. One of the foremost targets 
for high-precision studies is the electroweak sector that can be explored, for instance,
through the  multiple vector boson production channels. Among these, the production rate 
for a $Z$-boson and a photon is one of the highest and, at least for leptonic decays of the 
$Z$-boson, provides a final state whose particles can all be measured with excellent precision
\cite{Chatrchyan:2011rr,
	Aad:2011tc,
	Aad:2012mr,Chatrchyan:2013nda,Aad:2013izg,Chatrchyan:2013fya,%
	Aad:2014fha,
	Khachatryan:2015kea,Aad:2016sau,Khachatryan:2016yro,CMS:2016xzm,Aaboud:2017uhw,Aaboud:2016trl}.
It is important to
perform a careful comparison with the theoretical prediction of the \SM{} since
any deviation from it could indicate the presence of anomalous $Z\gamma\gamma$ or $ZZ\gamma$ 
couplings~\cite{Baur:1992cd}.  Beyond such tests, $Z\gamma$ production provides important 
backgrounds to many other searches. For instance, it is a crucial ingredient in background 
estimations for the rare $Z(\to l\bar l)\gamma$ Higgs decay \cite{Aaboud:2017uhw}. 
The kinematics of signal and background are very challenging in this case and require precise 
predictions to ensure an adequate understanding of both processes~\cite{Spira:1995rr,Djouadi:1996yq}. 
The $Z\gamma$ process also 
represents a leading background in searches for heavy resonances that decay into a $Z$-boson 
and a photon~\cite{Aad:2014fha,Aaboud:2017uhw,Aaboud:2016trl}, for example new singlet scalars with 
loop-induced 
decays~\cite{Low:2011gn}. Beyond that, the channel in which the $Z$ boson decays to neutrinos gives 
rise to events containing a photon and missing energy. This is therefore of great interest to 
searches for dark matter and more general \BSM{} 
studies with similar signatures~\cite{Fox:2011pm,Belanger:2012mk,Gabrielli:2014oya,Maltoni:2015twa}.

In order to provide the strongest point of comparison with the \SM{} it is therefore essential that 
the theoretical prediction for $Z\gamma$ production be available at the highest possible accuracy.  
The first steps towards providing a reliable calculation of this cross section, and related 
observables, were made by extending the initial \LO{} 
calculations~\cite{Brown:1979ux,Renard:1981es} to \NLO{} \QCD{}~\cite{Ohnemus:1992jn}. Further 
refinements were provided later on, to account for lepton decays in the narrow width approximation 
and to account for the effects of anomalous couplings~\cite{Ohnemus:1994qp,Baur:1997kz}. The 
calculation of one-loop helicity amplitudes for this process~\cite{Dixon:1998py} allowed the 
inclusion of such effects in a flexible Monte Carlo code~\cite{DeFlorian:2000sg} and in the 
general-purpose code \MCFM{}~\cite{Campbell:1999ah}.  The one-loop gluon fusion contribution, $gg 
\to Z\gamma$ formally enters at \NNLO{} \QCD{} but can also be computed 
separately~\cite{Ametller:1985di,vanderBij:1988fb,Adamson:2002rm} and is important for \LHC{} 
phenomenology~\cite{Campbell:2011bn}. The effects of \NLO{} electroweak (\EW{}) corrections have 
also been computed for an on-shell $Z$-boson~\cite{Hollik:2004tm,Accomando:2001fn,Accomando:2005ra} 
and have been recently included in a full \NLO{} \QCD{}+\EW{} prediction for this 
process~\cite{Denner:2014bna,Denner:2015fca}.  Finally, the first predictions for $Z\gamma$ 
production through \NNLO{} \QCD{} -- providing robust theoretical precision for this cross section 
and related quantities -- have been computed 
recently~\cite{Grazzini:2013bna,Grazzini:2015nwa}.

In this publication we provide an independent calculation of $Z\gamma$ production at \NNLO{} in 
\QCD{}. Given the importance of this process, an independent verification such as the one we 
provide here is invaluable.\footnote{For example in the case of \NNLO{} $\gamma\gamma$ production
the first published results \cite{PhysRevLett.108.072001} were not confirmed by an independent
calculation \cite{Campbell:2016yrh}, but have since been updated \cite{PhysRevLett.117.089901}.  }
We extend the previously available results to also 
include the case of anomalous $Z\gamma\gamma$ and $ZZ\gamma$ couplings at \NNLO{}. This allows them 
to be probed with greater confidence and enables more accurate limits to be placed on their possible 
values. Note that we use the label $Z\gamma$ here and in the following as a shorthand for calculations that
represent the final states with charged leptons $l^+l^-\gamma$ and neutrinos $\nu\bar\nu\gamma$.
In particular, in the former case we include the effects of photon radiation in the $Z$ boson decay and
also virtual photon contributions.

Our calculation is based on matrix elements that, for numerical efficiency, have been computed 
analytically and that are combined to provide a full \NNLO{} calculation using the jettiness phase 
space slicing method~\cite{Boughezal:2015dva,Gaunt:2015pea}. This type of approach to the 
computation of \NNLO{} corrections is in common with the previous 
calculation~\cite{Grazzini:2013bna,Grazzini:2015nwa}, that employed a slicing method known as 
$q_T$-subtraction~\cite{Catani:2007vq}. These techniques rely on the choice of a slicing parameter 
that is sufficiently small that either the associated power corrections are negligible or they can 
be extrapolated away. We perform a detailed check of the jettiness slicing procedure, which is 
especially important since the numerical uncertainties achieved with slicing methods can be of a 
similar size as the scale uncertainty observed at \NNLO{}. Focusing on the technical aspects of our 
calculation is important to validate our results so that they can be made publicly available in the 
next release of \MCFM{}. 

In \cref{sec:setup} we describe the ingredients of our calculation and how we obtain the 
results presented in the rest of the paper. \Cref{sec:comparison} provides a detailed 
comparison with the results previously obtained in ref.~\cite{Grazzini:2015nwa}. With our method validated, we 
proceed to a discussion of \SM{} phenomenology in \cref{sec:SMpheno}, discussing both 
$Z$-boson 
decays into neutrinos and into charged leptons as an application for the Higgs production 
background. In \cref{sec:anomcoup} we assess the impact of \NNLO{} corrections on the extraction of 
anomalous coupling limits. Finally, our conclusions and outlook are given in \cref{sec:conc}.

\section{Calculation and setup}
\label{sec:setup}

Computing $Z\gamma$ to \NNLO{} accuracy requires the calculation and assembly of numerous 
contributions that we list here.  The existing 0-jettiness slicing implementation in
\MCFM{}~\cite{Boughezal:2016wmq}, reviewed briefly below, is used to easily combine all the
necessary amplitudes  to produce a full \NNLO{} code.

We have taken the $Z\gamma$ 0/1/2-loop amplitudes from ref.~\cite{Gehrmann:2011ab}, 
using the function \texttt{tdhpl} \cite{Gehrmann:2001jv} to evaluate the two dimensional harmonic 
polylogarithms through which they are expressed.  The 
double radiation tree amplitudes and the one-loop single radiation amplitudes are based on the 
previous implementations in \MCFM{} \cite{Campbell:2012ft} (see also \cite{Bern:1997sc}), but are 
completely rewritten for readability and run-time performance.\footnote{We have fixed a bug 
affecting the radiation in decay amplitudes in the existing $Z\gamma j$ implementation in 
MCFM~\cite{Campbell:2012ft}.}
 For the charged lepton decay channel the radiation of 
a photon from the leptons is possible, which requires the evaluation of the two-loop quark 
formfactor. We have taken it from ref.~\cite{Gehrmann:2010ue}, where it is given up to three loops, 
see also \cite{Baikov:2009bg,Lee:2010cga}. Note that the two-loop result has previously already 
been published in ref.~\cite{Matsuura:1988sm}.

We have checked that our tree level and one-loop matrix elements agree with results obtained from 
\OL{} \cite{Cascioli:2011va}. Additionally, we found agreement with \MG{} 4.2.7
\cite{Alwall:2007st} for our one-loop real emission matrix elements. We work
consistently in the five flavor scheme with a zero bottom quark mass.

The current release version 8.0 of \MCFM{} includes $Z\gamma$ amplitudes with anomalous couplings 
at \NLO{}
\cite{DeFlorian:2000sg}. At \NNLO{} one requires the double virtual corrections, easily obtained
from the two-loop quark form factor.
We additionally implemented the double real radiation anomalous coupling tree 
amplitudes using the vertices from ref.~\cite{Hagiwara:1986vm} (following the convention with an 
additional factor of $i$ as 
in ref.~\cite{Gounaris:1999kf}). Specifically we contracted the vertices with the four parton plus 
$Z$ boson off-shell current \cite{Berends:1988yn,Bern:2004ba}\footnote{The hadronic $Z$ current 
$S(+;-:+;-)_{\dot{A}B}$ given in the appendix of ref.~\cite{Berends:1988yn} for the $q\bar{q}gg$ 
case has a typo. The current given in ref.~\cite{Giele:1991vf} agrees with our own calculation.} to 
obtain the complete amplitudes. To obtain the real emission one-loop amplitudes we contracted the 
anomalous coupling vertices with the $V\to 3j$ loop amplitude current from 
ref.~\cite{Garland:2002ak} and performed the analytic continuations to $2\to2$ kinematics as 
outlined in ref.~\cite{Gehrmann:2002zr}.

\paragraph{Jettiness subtractions.}

Since the assembly of our calculation relies on the method of jettiness subtractions, which 
has been implemented in \MCFMEIGHT{} for color singlet final states \cite{Boughezal:2016wmq}, we 
give a brief overview of this method here. For a more complete review we refer the reader to 
the referenced literature.
The $N$-jettiness subtraction formalism is a phase space slicing scheme 
for combining infrared singular matrix elements entering higher order perturbative QCD calculations 
that contain $N$ colored final state partons at Born level \cite{Boughezal:2015dva,Gaunt:2015pea}. 
Using the event shape variable $N$-jettiness \cite{Stewart:2010tn} , $\tau^N$, the formalism 
allows for handling processes with initial- and final-state infrared singularities and any number of
colored final state particles.
The calculation of the cross section proceeds, not fundamentally different from classic \NLO{} 
slicing schemes, through a separation of the infrared and collinear singular phase space using 
a jettiness cutoff parameter $\tau_\text{cut}^N$.
The below-cut part is predicted by soft collinear effective 
theory (\SCET{}) in terms of a factorization theorem using the hard scattering matrix element and 
soft and beam functions.\footnote{We note that factorization theorems in \SCET{} 
with respect to other variables also constitute a powerful method for performing higher order 
calculations; for example top-quark decay has been calculated in such a way at \NNLO{} 
\cite{Gao:2012ja}.} 
Here, for $Z\gamma$ production, we can restrict 
ourselves to the case of $N=0$ colored final state partons at 
Born level and refer to $\tau\equiv\tau^{N=0}$ in the following. For a parton scattering event 
$p_a + p_b \to X + p_1 + \ldots + p_M$ with parton momenta $p_a,p_b,p_1,\ldots,p_M$ and color 
singlets $X$, the event shape variable $0$-jettiness is defined by
\[
	\tau = \sum_{k=1}^M \min_{i=a,b} \left\{ \frac{2 q_i p_k}{Q_i} \right\}\,,
\]
where $q_i$ are massless Born reference momenta defining the jettiness axes and $Q_i$ are 
normalization factors. For a \NNLO{} cross section, by demanding $\tau<\tau_\text{cut}$ the doubly 
unresolved region can be 
isolated and the real emission matrix elements in that region can be integrated over analytically 
in the soft/collinear approximation and finally added to the virtual contributions. The part 
$\tau>\tau_\text{cut}$ corresponds to the singly unresolved limit of the \NLO{} process with an 
additional jet.  It can be handled by any \NLO{} subtraction scheme, preferably using a fully local 
\NLO{} subtraction scheme like Catani-Seymour subtractions \cite{Catani:1996vz,Catani:1996jh}.
The sum of the two contributions yields a result that, in the limit that $\tau_\text{cut}\to0$,
returns the complete NNLO cross section.

\paragraph{Choice of parameters.} \label{par:parameters}

Our electroweak couplings are derived from the input parameters $m_W=\SI{80.385}{GeV}$, 
$m_Z=\SI{91.1876}{GeV}$ and $G_F={1.6639\cdot10^{-5}}{\text{GeV}^{-2}}$; additionally we have 
$\Gamma_Z=\SI{2.4952}{GeV}$ \cite{Olive:2016xmw}. We use \MMHT{} 2014 \PDF{}s 
\cite{Harland-Lang:2014zoa} with $\alpha_s$ values provided by the \PDF{} set at the corresponding loop 
orders.

All our cross sections use a common central renormalization and factorization scale of $\mu_0 =\sqrt{m_V^2 + (p_T^\gamma)^2}$. For scale uncertainties we use a seven point variation
in which antipodal variations of $\mu_R$ and $\mu_F$ are excluded.
Specifically, we take the maximum and minimum
values resulting from the use of scales in the set
$$ (\mu_R,\mu_F)\in\{(1,\nicefrac{1}{2}), (1,2), (\nicefrac{1}{2},1), (2,1), (\nicefrac{1}{2},\nicefrac{1}{2}), (2,2), (1,1) \}\cdot \mu_0\,. $$

Unless specified otherwise for comparison reasons, we use the boosted jettiness
definition \cite{Moult:2016fqy} which is defined in the color singlet center of
mass frame, rather than the hadronic definition that is used in \MCFMEIGHT{} and
defined in the hadronic center of mass frame. See
refs.~\cite{Stewart:2010tn,Stewart:2009yx,Stewart:2010pd,Gaunt:2015pea,Berger:2010xi}
for discussion.

To avoid infrared singularities arising from the emission of photons from partons
we use the Frixione smooth cone isolation 
criterion \cite{Frixione:1998jh}.  In this method one defines a cone of radius $R=\sqrt{(\Delta\eta)^2 + (\Delta\phi)^2}$
around the photon where $\Delta\eta$ and $\Delta\phi$ are the pseudorapidity and azimuthal angle 
difference between the photon and any parton. The total partonic transverse energy inside
a cone with radius $R$ is then required to be smaller than 
\[
	\epsilon \cdot E_\gamma \left( \frac{1-\cos R}{1-\cos R_0} \right)^n\,,
\]
for all cones $R<R_0$, where $E_\gamma$ is the transverse photon energy, and $\epsilon$, $R_0$
and $n$ are parameters.
When comparing with the results of ref.~\cite{Grazzini:2015nwa}
we adopt their choice of the photon isolation parameters, $\epsilon=0.5,\,n=1$ and $R_0=0.4$. Elsewhere 
we use the isolation parameters $\epsilon=0.1,\,n=2$ and $R_0=0.4$ that for isolated 
prompt photon production at \NLO{} provide results similar to those obtained using fragmentation
functions and experimental cuts \cite{Campbell:2016yrh,Campbell:2016lzl}.

\paragraph{Slicing cut-off parameter extrapolation.}

The leading behavior of the \NLO{} and \NNLO{} cross section coefficients 
$\Delta\sigma_\text{NLO}$, $\Delta\sigma_\text{NNLO}$ for $\taucut\to0$ is given by,
\begin{equation}
\Delta\sigma_\text{NLO}^\text{asympt.} = \left(\Delta\sigma_\text{NLO} + 
c_1\cdot\frac{\taucut}{Q}\log\frac{\taucut}{Q} \right) + \ldots 
\label{eq:nloasymptote}
\end{equation}
\begin{equation}
\Delta\sigma_\text{NNLO}^\text{asympt.} = \left(\Delta\sigma_\text{NNLO} 
+ c_3\cdot\frac{\taucut}{Q}\log^3\frac{\taucut}{Q} \right) + \ldots \,.
\label{eq:deltannloasymptote}
\end{equation}
It is these forms that we will use for fitting. We will treat the scale $Q$ as well 
as the coefficients $c_i$ as nuisance parameters that are fitted with the data but unused. 
See for example ref.~\cite{Moult:2016fqy} for a more detailed discussion of the higher power 
corrections and their fitting.

In all cases we weight the data points in the non-linear fitting procedure with the inverse square 
of their Monte Carlo integration numerical uncertainty. Our quoted uncertainties for results
that have been extrapolated in the limit $\taucut\to0$, $\Delta\sigma_\text{NLO}$ and $\Delta\sigma_\text{NNLO}$,
correspond to a $95\%$ confidence level. Including higher power corrections is equivalent to a 
theoretical uncertainty in the fit model. To be of use for the prediction, these additional 
nuisance parameters would require much smaller statistical uncertainties than we demand for the
phenomenology in this study. They would become important when their nominal contribution 
becomes comparable with the statistical uncertainties in our predictions \cite{Moult:2016fqy}. In any case 
we make sure that within our uncertainties the results are unchanged by excluding data points with 
our largest and smallest considered $\taucut$ values. We discuss the extrapolation in more detail 
in \cref{sec:Grazzinicomparison} using actual examples.

\subsection{Improvements to jettiness slicing}

Before we present our main results in the following sections, we first investigate potential improvements to our implementation 
of jettiness slicing from two sources. Since these improvements are best studied in light of a
cutoff independent calculation, we compare \NLO{} results to those obtained by Catani-Seymour subtractions.
The first improvement may be obtained by including the effects of subleading power  
corrections~\cite{Moult:2016fqy,Boughezal:2016zws} for Drell-Yan type color-singlet production. The 
second improvement results from a boosted definition of the jettiness slicing variable.

\begin{figure}
	\includegraphics[width=\columnwidth]{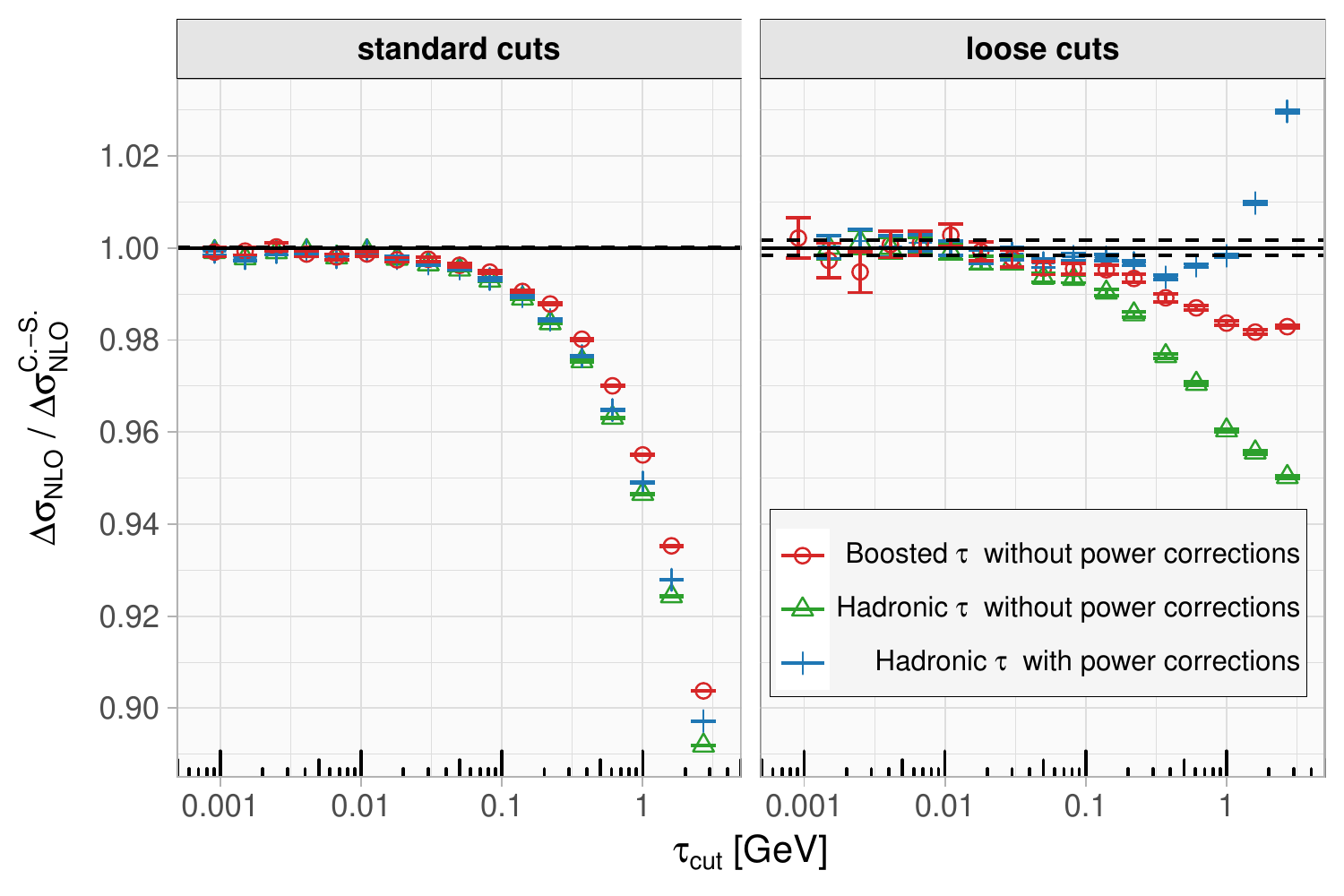}
	\caption{Normalized \NLO{} cross section coefficient for the charged lepton decay channel with 
	standard 
	cuts given in \cref{tab:cuts-8TeV-Zepem} and loose cut modifications in \cref{eq:loosecuts}. 
	Shown is the dependence on the jettiness slicing cutoff parameter $\tau_\text{cut}$.
	The results are obtained using the hadronic (\MCFMEIGHT{}) jettiness definition with and 
	without power corrections and additionally for the boosted definition. For performance 
	comparison, the cross section is normalized to the cutoff independent Catani-Seymour result. 
} 
	\label{fig:jettiness-improvements}
\end{figure}

We consider the case of a $Z$-boson decaying into charged leptons at $\sqrt{s}=\SI{8}{\TeV}$
using the cuts shown in \cref{tab:cuts-8TeV-Zepem}.   Fig.~\ref{fig:jettiness-improvements} (left)
shows the approach of the jettiness calculation of the 
\NLO{}  coefficient to the result obtained using Catani-Seymour (C.-S.) subtraction 
\cite{Catani:1996vz,Catani:1996jh}, both with 
and without the inclusion of the dominant subleading power corrections for the hadronic jettiness 
definition used in \MCFMEIGHT{}. We see that the inclusion 
of the subleading power corrections has almost no effect. We surmise that this is due to the fact 
that, in 
contrast to the processes studied in refs.~\cite{Moult:2016fqy,Boughezal:2016zws}, $Z\gamma$ 
production contains $t$-channel diagrams at Born level. To investigate this issue further, we repeat this analysis under a set 
of cuts that decreases the importance of such diagrams relative to Drell-Yan type $s$-channel 
contributions in 
which the photon is radiated from a charged lepton. This set of cuts corresponds to those specified 
in \cref{tab:cuts-8TeV-Zepem}, with the replacements:
\begin{equation}
p_\text{T}^l > \SI{20}{\GeV}\,, \quad p_\text{T}^\gamma > \SI{10}{\GeV} \,, \quad \Delta R(l,\gamma) > 0.1 \,.
\label{eq:loosecuts}
\end{equation}

\begin{table}[]
	\centering
	\caption{Applied cuts for $Z\to e^+e^-$ decay at a center of mass energy 
		$\sqrt{s}=\SI{8}{\TeV}$.}
	\vspace*{1em}
	\begin{tabular}{l|c}
		Leptons & $p_\text{T}^l > \SI{25}{\GeV}$, $\abs{\eta^l} < 2.47$ \\
		\multirow{2}{*}{Photon} & $p_\text{T}^\gamma > \SI{40}{\GeV}$, $\abs{\eta^\gamma}<2.37$ \\
		& Frixione isolation $\epsilon_\gamma=0.5, R_0=0.4, n=1$ \\
		Jets & anti-$k_\text{T}$, $D=0.4$, $p_\text{T}^\text{jet} > \SI{30}{\GeV}$, 
		$\abs{\eta^\text{jet}}<4.5$ \\
		Separation & $m_{l^+l^-}>\SI{40}{\GeV}$, $\Delta R(l,\gamma) > 0.7$, $\Delta 
		R(l/\gamma,\text{jet}) > 0.3$
	\end{tabular}
	\label{tab:cuts-8TeV-Zepem}
\end{table}

The results of this study are shown in Fig.~\ref{fig:jettiness-improvements} (right). Without power 
corrections the approach to the correct result is similar to the approach when using our 
standard cuts that is shown in the left panel. However, the inclusion of power corrections now makes
a much more marked effect and significantly increases the value of $\taucut$ that might be 
considered asymptotic. We therefore conclude that the effect of subleading power corrections is 
process-specific and that the inclusion of the results of 
refs.~\cite{Moult:2016fqy,Boughezal:2016zws} provides no  significant benefit under the cuts  
considered here. Therefore, and for consistency, we do not include them in any of our subsequent 
results.

We now compare the performance of the jettiness slicing procedure with two different definitions of 
$\taucut$.  These correspond to the version in the hadronic center of mass frame employed
for other color-singlet processes in \MCFMEIGHT{} and the boosted one in the color 
singlet center of mass frame. The results of this comparison
are also shown in \cref{fig:jettiness-improvements}.
We see that the improvement when using the boosted definition is  relatively small for the standard cuts, 
due to the fact that our set of cuts demand the production of a $Z\gamma$ system that has a rather 
high virtuality and is quite central. Despite this, the small improvement in the approach to the 
asymptotic value is the result of a trivial change in the code. We therefore adopt the boosted 
definition for all the studies presented in this paper.

\begin{figure}
	\includegraphics[width=\columnwidth]{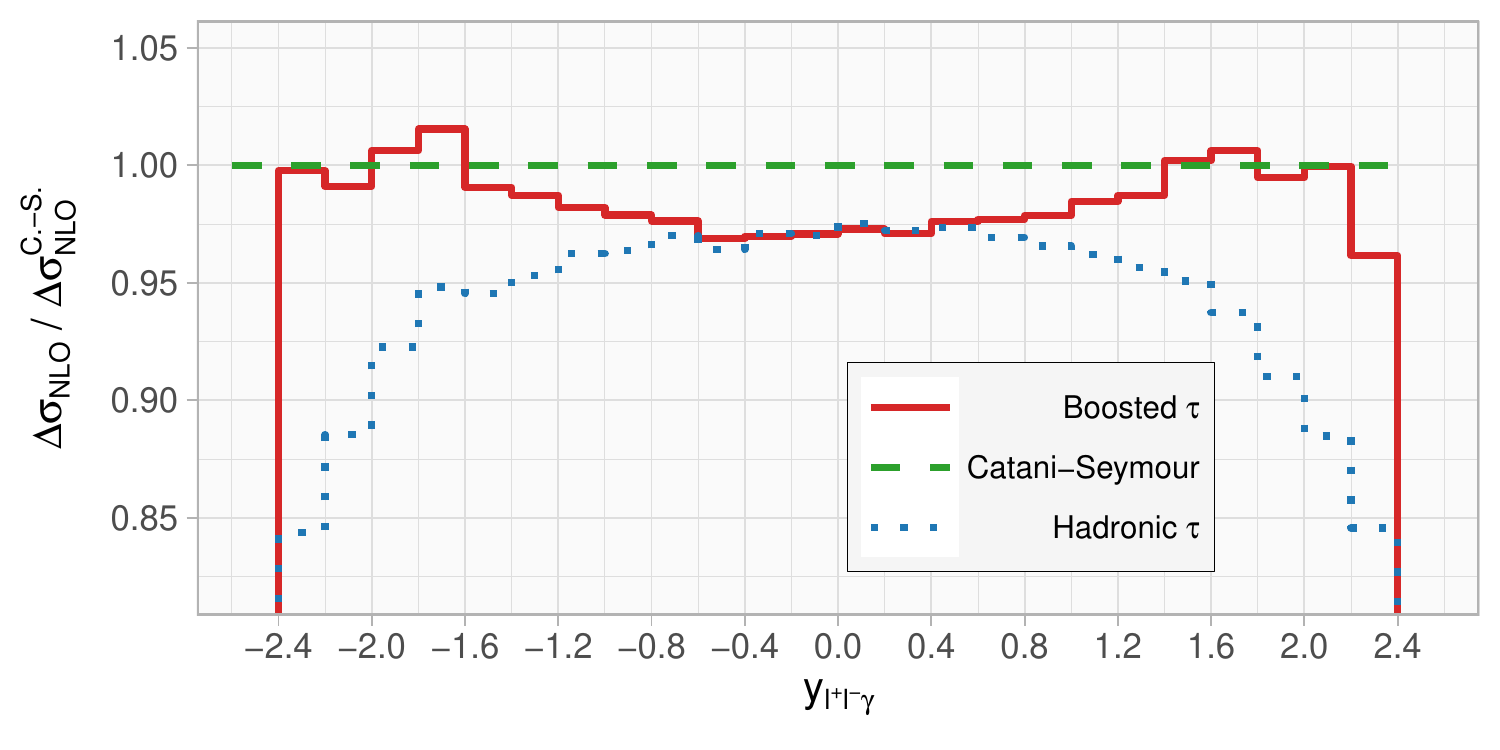}
	\caption{Normalized \NLO{} coefficient pseudorapidity distribution for the charged lepton decay 
	channel with standard cuts given in \cref{tab:cuts-8TeV-Zepem} and loose cut modifications in 
	\cref{eq:loosecuts}. The distributions are obtained with $\tau_\text{cut}=\SI{1.6}{\GeV}$ using 
	the boosted and hadronic (\MCFMEIGHT{}) jettiness definitions, normalized to the Catani-Seymour 
	obtained result. The numerical uncertainties of the outermost bins are large since the 
	distribution has been generated on the fly for the integrated cross section. }
	\label{fig:rapidity-boosted-vs-hadronic}
\end{figure}

For a more refined comparison it is instructive to consider not just a total cross section but a differential
prediction. \Cref{fig:rapidity-boosted-vs-hadronic} shows
the pseudorapidity distributions of the \NLO{} coefficients computed using our loose cuts with 
Catani-Seymour 
subtraction and using jettiness slicing with $\tau_\text{cut}=\SI{1.6}{\GeV}$ for both boosted and 
hadronic definitions. For the integrated cross section, as can 
be seen in \cref{fig:jettiness-improvements}, the difference between the boosted and hadronic 
definition is about $2.5\%$. Nevertheless, the difference for rapidities greater than about 
one already exceeds $2.5\%$ and grows to about $10\%$ for rapidities larger than two. The 
difference of the integrated cross section between using C.-S. and boosted jettiness is about two 
percent that manifests as a relatively constant difference throughout the whole 
rapidity range. This clearly shows that when computing arbitrary differential distributions one cannot
generally rely on the smallness of the residual $\tau_\text{cut}$ dependence of the integrated 
cross section, but one has to consider it differentially as well. For the differential distributions 
considered in our study we are not affected by this problem, as we show in \cref{sec:SMpheno}. This 
is partly due to directly using the boosted definition, and additionally to just using 
asymptotically small enough $\tau_\text{cut}$ values where the numerical integration uncertainty 
dominates over the residual $\tau_\text{cut}$ error. 

\section{Validation} \label{sec:Grazzinicomparison}
\label{sec:comparison}

The process of $Z\gamma$ production has been considered previously at 
\NNLO{} in refs.~\cite{Grazzini:2013bna,Grazzini:2015nwa} in the $q_T$ subtraction formalism 
\cite{Catani:2007vq}. For comparison and mutual validation we check some results of
ref.~\cite{Grazzini:2015nwa} in this section. The \SI{8}{TeV} cuts used in this comparison have 
already been summarized in \cref{tab:cuts-8TeV-Zepem} for the  charged lepton decay channel;  the 
counterparts for the neutrino decay channel are shown in \cref{tab:cuts-8TeV-Znunu}. 

Specifically we will compare our results with those provided in tables~3
and~5 of ref.~\cite{Grazzini:2015nwa}, for the \SI{8}{\TeV} jet inclusive
case.  Their stated conservatively-estimated \NNLO{} numerical uncertainty is
between $0.5\%$ and $0.6\%$. Note that the authors do not give a technical
explanation of how their extrapolation is performed and the uncertainty is
estimated, but only state that it is a combination of statistical Monte Carlo
integration error and an extrapolation procedure. As such, our results can
evaluate whether their uncertainty estimation is robust.

In order to minimize the uncertainty resulting from the jettiness slicing method
we compute the \NNLO{} cross section coefficient separately from the \NLO{} cross 
section, which is obtained by Catani-Seymour dipole subtractions. From the quoted \NNLO{} results in 
ref.~\cite{Grazzini:2015nwa} we subtract our \NLO{} results obtained with \NNLO{} \PDF{}s in
order to translate them into an expectation for the \NNLO{} coefficient. Since our \NLO{} results
mutually agree within $0.2\%$, we consider such a procedure for extracting the \NNLO{} coefficients 
to be reasonable.

\begin{table}[]
	\centering
	\caption{Applied cuts for $Z\to \bar\nu\nu$ decay at a center of mass energy $\sqrt{s}=\SI{8}{\TeV}$.}
	\vspace*{1em}
	\begin{tabular}{l|c}
		Neutrinos & $p_\text{T}^{\bar\nu\nu} > \SI{100}{\GeV}$ \\
		\multirow{2}{*}{Photon} & $p_\text{T}^\gamma > \SI{130}{\GeV}$, $\abs{\eta^\gamma}<2.37$ \\
		& Frixione isolation $\epsilon_\gamma=0.5, R_0=0.4, n=1$ \\
		Jets & anti-$k_\text{T}$, $D=0.4$, $p_\text{T}^\text{jet} > \SI{30}{\GeV}$, $\abs{\eta^\text{jet}}<4.5$ \\
		Separation & $\Delta R(\gamma,\text{jet}) > 0.3$
	\end{tabular}
	\label{tab:cuts-8TeV-Znunu}
\end{table}

We begin with a comparison in the neutrino decay channel, where the authors of ref.~\cite{Grazzini:2015nwa}
report cross  sections of \SI{42.33}{fb}, \SI{70.98}{fb} and \SI{80.82}{fb} at \LO{}, \NLO{} and \NNLO{}, 
respectively.  Their stated uncertainty estimate is 0.5\% for the \NNLO{} result from integration 
error and finite $q_T$ cut, which translates to \SI{0.4}{fb} in absolute terms. We assume that 
their uncertainty for the \LO{} and \NLO{} results corresponds to the stated number of significant 
figures, translating to an uncertainty of \SI{0.01}{fb}.
Our \LO{} cross section of \SI{42.35}{fb} is in good agreement to better than half a per-mille with 
their result.

\begin{figure}
	\includegraphics[width=\columnwidth]{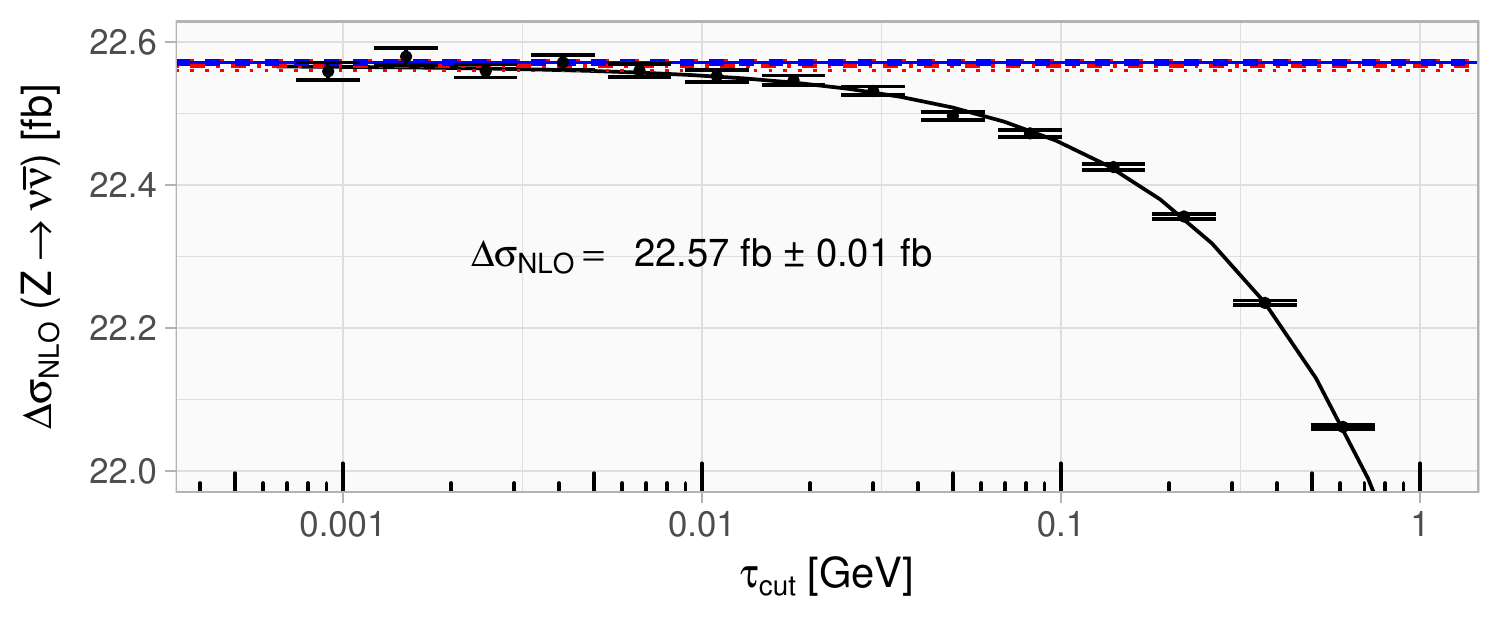}
	\caption{$\taucut$ dependence of the $Z(\to \nu\bar\nu)\gamma$ \NLO{} cross section coefficient 
	with \NLO{} \PDF{}s and cuts given in \cref{tab:cuts-8TeV-Znunu}. The horizontal blue solid and 
	dashed lines represent the C.-S. obtained 
	result of \SI{22.572(3)}{fb} and its numerical uncertainty, respectively. The horizontal red 
	dot-dashed and dotted lines represent the fitted 
	asymptotic value of \SI{22.57(1)}{fb} and its numerical uncertainty, respectively, while the 
	solid black 
	line shows the fitted result.}
	\label{fig:8TeV-Znunu-taucut-nlo}
\end{figure}

To get a first estimate of how low $\tau_\text{cut}$ should be in order to reach asymptotic 
behavior for the $\taucut\to0$ extrapolation, we present in \cref{fig:8TeV-Znunu-taucut-nlo} the 
$\taucut$ dependence of the 
\NLO{} result computed with jettiness slicing.  As before, the results are compared 
with the one obtained by using Catani-Seymour dipole subtraction that has already been implemented in \MCFM{}.
For the fitting to the asymptotic formula in \cref{eq:nloasymptote} we have used all data points. We have 
also limited the fitting data to sets of $\taucut<\SI{0.5}{\GeV}$ and $\taucut<\SI{0.08}{\GeV}$. In both cases the fitted 
value of $\Delta\sigma_\text{NLO}$ and its uncertainty stay the same, while of course increasing the fit 
uncertainty of the nuisance parameters affecting the tail. It is also possible to remove some of 
the smallest $\taucut$ data points 
without altering the asymptotic value and uncertainty, to some extent. This makes us believe that 
the dominant behavior to extract $\Delta\sigma_\text{NLO}$ is captured around the region 
$\taucut\simeq \SI{0.01}{\GeV}$.

Since our cross section uncertainties are not small enough to significantly fix the values of the 
nuisance parameters $c_1$ and $Q$ in \cref{eq:nloasymptote}, we do not include higher order power 
corrections. Including them deteriorates the fit quality of the nuisance parameters drastically. 
This follows the guidance in ref.~\cite{Moult:2016fqy}: in order to include additional terms in 
the fit, the 
model uncertainty induced by missing higher order power corrections in the fit should be comparable 
to the statistical uncertainties of the cross section values. For example the nuisance parameters 
in \cref{fig:8TeV-Znunu-taucut-nlo} are still fitted with standard uncertainties of 50--60\%.

Adding the \LO{} contribution to the \NLO{} coefficient, our result of \SI{71.09(1)}{fb} compares 
well with the result from 
ref.~\cite{Grazzini:2015nwa} of \SI{70.98}{fb}. They agree within $0.2\%$, but leave open 
the question of a possible tiny systematic difference in, for example, phase-space sampling.
Our code for the \NLO{} cross section with default settings is stable to values of $\taucut$ 
at least as small as $\SI{e-4}{\GeV}$, but at that point the numerical uncertainty using the same 
computational runtime increases drastically, of course.  Instead we are able to obtain high precision 
results 
by sampling multiple larger values of $\taucut$ using the additional information inferred from the 
asymptotic behavior in \cref{eq:nloasymptote}. This also makes the result robust against 
statistical noise, underestimated uncertainties, or a possibly biased Monte Carlo grid, since our 
results for different values of
$\tau_\text{cut}$ are statistically independent.
Note that even at $\taucut=\SI{0.02}{\GeV}$ the deviation of the \NLO{} coefficient from its asymptotic value 
is just about a percent. Using a value of $\taucut\lesssim\SI{0.02}{\GeV}$ for the calculation of the \NNLO{} 
coefficient is then sufficient for any phenomenological application considering its 
relative contribution to the \NLO{} cross section, as we will see.

\begin{figure}
	\includegraphics[width=\columnwidth]{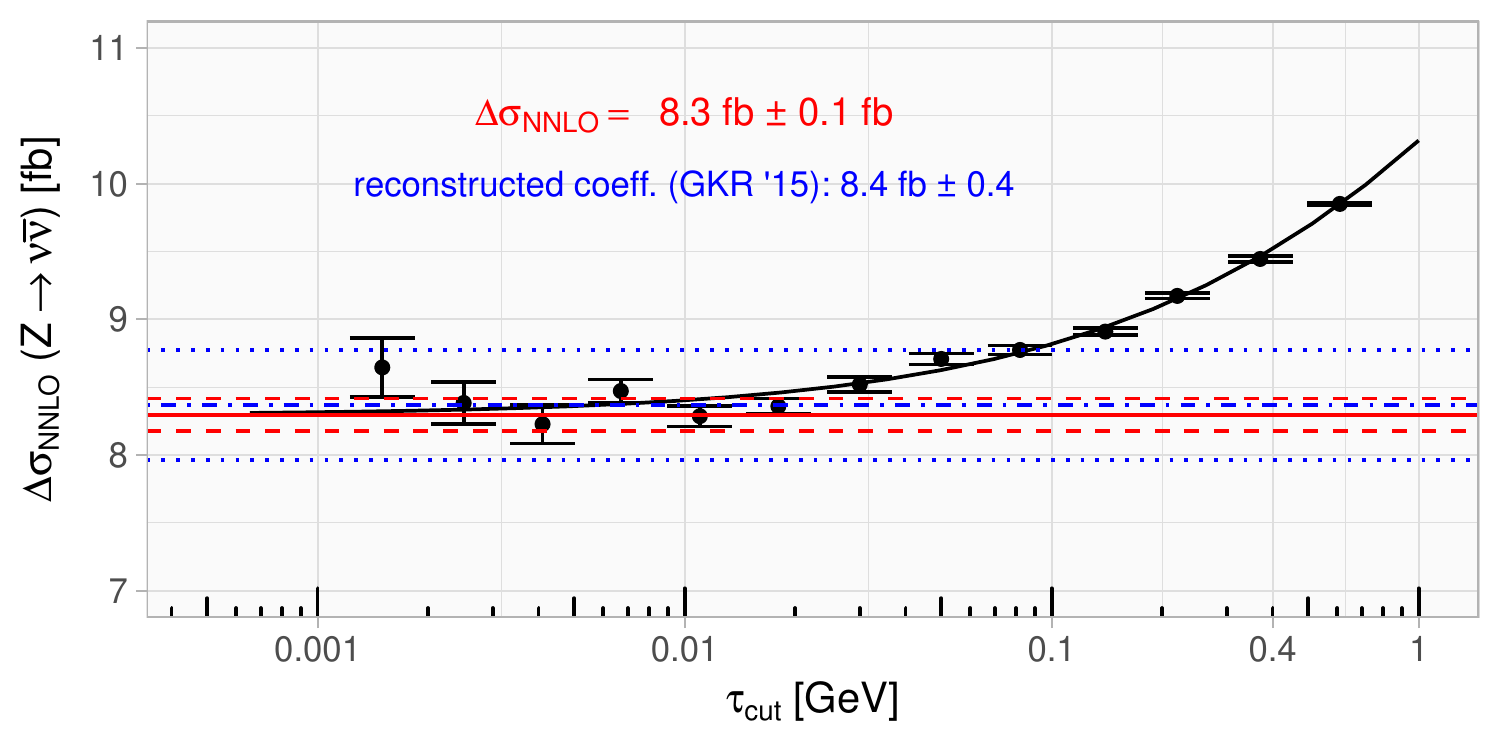}
	\caption{$\taucut$ dependence of \NNLO{} cross section coefficient in the neutrino decay 
	channel with given cuts in \cref{tab:cuts-8TeV-Znunu} . The horizontal red solid and dashed 
	lines represent the fitted asymptotic value of \SI{8.3\pm0.1}{fb} and its numerical 
	uncertainty, respectively, while the solid black line shows the fitted result. The blue 
	dot-dashed 
	and dotted lines show the reconstructed \NNLO{} coefficient and its numerical uncertainty, 
	respectively, from the \NNLO{} result given in 
	ref.~\cite{Grazzini:2015nwa}.}
	\label{fig:8TeV-Znunu-taucut-nnlocoeff}
\end{figure}

We show the $\tau_\text{cut}$ dependence of the \NNLO{} cross section coefficient in 
\cref{fig:8TeV-Znunu-taucut-nnlocoeff}. The \NNLO{} coefficient extracted from the asymptotic fit 
contributes  \SI{8.3\pm0.1}{fb} to the full \NNLO{} result of \SI{80.7\pm0.1}{fb}. Our result 
is in excellent agreement with the result of \SI{80.8\pm0.4}{fb} given in 
ref.~\cite{Grazzini:2015nwa}.
Let us emphasize that our quoted uncertainty is purely statistical, originating from the Monte Carlo 
integration uncertainty in the weighted non-linear fit to the asymptotic $\taucut$ behavior. The 
uncertainty induced by the actual fitting model is subleading.

\begin{figure}
	\includegraphics[width=\columnwidth]{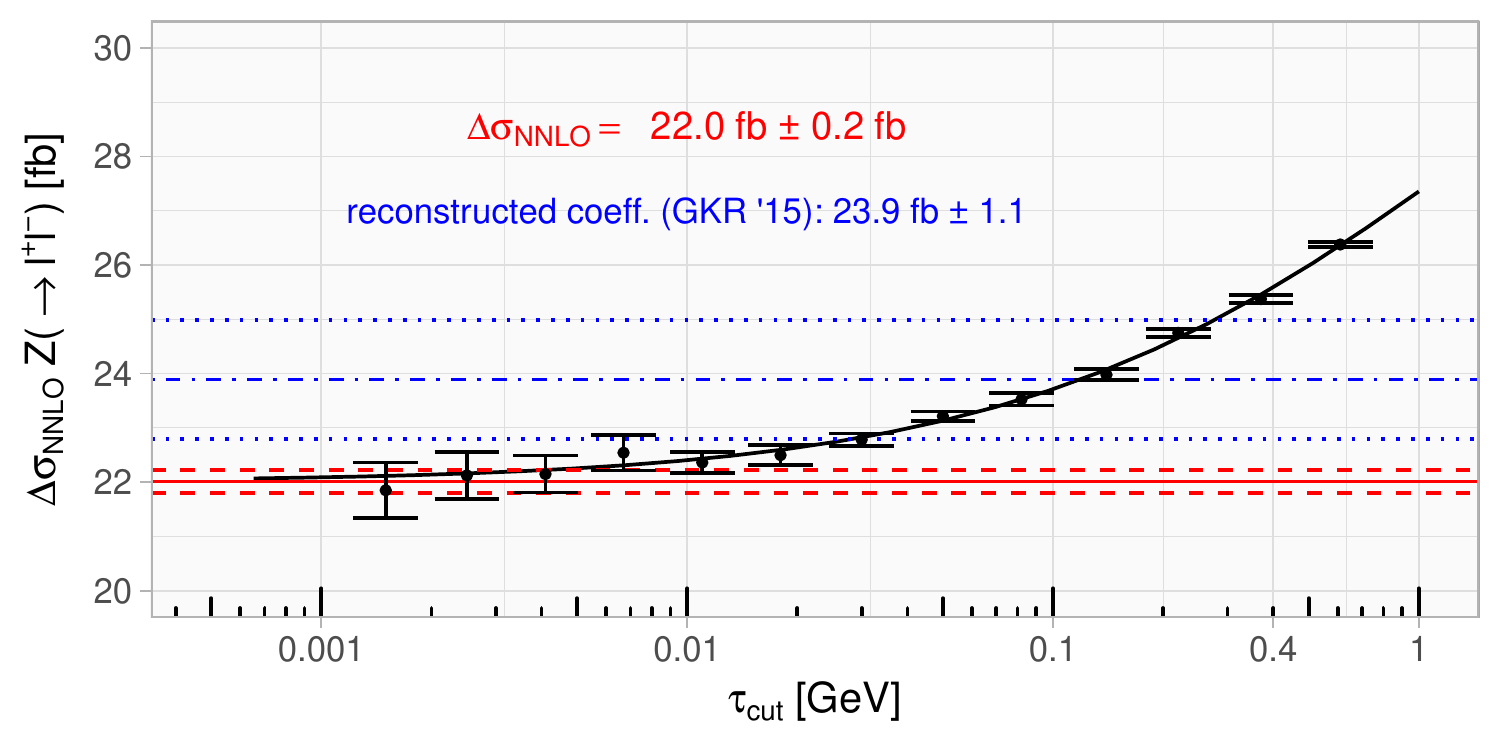}
	\caption{$\taucut$ dependence of the \NNLO{} cross section coefficient in the charged lepton 
	decay channel with cuts given in \cref{tab:cuts-8TeV-Zepem}. The horizontal red solid and 
	dashed lines represent the fitted asymptotic value of \SI{22.0\pm0.2}{fb} and its numerical 
	uncertainty, respectively, while the solid 
	black line shows the fitted result. The blue dot-dashed and dotted lines show the reconstructed 
	\NNLO{} coefficient and its numerical uncertainty, respectively, from the \NNLO{} result given 
	in ref.~\cite{Grazzini:2015nwa}.  }
	\label{fig:8TeV-Zepem-taucut-nnlocoeff}
\end{figure}

For the charged lepton decay channel our \LO{} result of \SI{84.115(5)}{fb} agrees with the result 
of \SI{84.09}{fb} in table~3 of ref.~\cite{Grazzini:2015nwa} within $0.03\%$. 
Our \NLO{} coefficient obtained from $\tau_\text{cut}\to0$ extrapolation is 
\SI{53.89(4)}{fb} and agrees perfectly with the C.-S. result of \SI{53.92(1)}{fb} within 
half a per-mille. This results in a total \NLO{} cross section of \SI{153.15(2)}{fb} that compares 
well with \SI{153.1}{fb} given in table~3 of \cite{Grazzini:2015nwa}.
Using the fitted \NNLO{} coefficient shown in 
\cref{fig:8TeV-Zepem-taucut-nnlocoeff} of \SI{22.0\pm0.2}{fb} we obtain a total \NNLO{} cross 
section of \SI{178.2\pm0.2}{fb}, with a relative uncertainty of about one per-mille. We compare this 
to the value
of \SI[separate-uncertainty = true]{180\pm1}{fb} in ref.~\cite{Grazzini:2015nwa} and find only 
broad compatibility. We note that their uncertainty is considerably larger than ours.
 This does not indicate a definite discrepancy but it is possible that the choice
 of slicing parameter could lead to such a difference. We observe that if we had instead chosen
 a larger value of $\tau_\text{cut}\sim\SI{0.1}{\GeV}$ then our results for both decay modes
 would have been compatible with those of ref.~\cite{Grazzini:2015nwa}, within their quoted uncertainties.

\section{Standard Model phenomenology at \SI{13}{TeV}}
\label{sec:SMpheno}

In order to present new phenomenological results and to study how the extrapolation
of $\tau_\text{cut}\to0$  performs at a higher center 
of mass energy, we calculate cross sections 
and kinematical distributions for the neutrino decay channel at $\sqrt{s}=\SI{13}{\TeV}$ in the 
following \cref{sec:Znunu}. 
Note again that we use different photon isolation cuts here compared to the \SI{8}{TeV} 
comparison results in \cref{sec:Grazzinicomparison}; specifically we set $\epsilon=0.1,\,n=2$
and $R_0=0.4$ (see \cref{par:parameters}). In 
\cref{sec:Hbackground} we study the \SI{13}{\TeV} $m_{l\bar l \gamma}$ spectrum as a 
background to the very rare $H\to Z(\to l\bar l)\gamma$ decay.

\subsection{Decay to neutrinos} \label{sec:Znunu}

We expect that the cuts for the upcoming \SI{13}{TeV} $Z\gamma$ \ATLAS{} analysis will be very similar
to the ones that are summarized in \cref{tab:cuts-13TeV-Znunu}. Using these we investigate \NNLO{} predictions for the cross section and
kinematical distributions, where in the latter case we discuss the systematic uncertainty induced by 
using a finite value of $\tau_\text{cut}$ instead of performing a systematic extrapolation 
procedure.

\begin{table}[]
	\centering
	\caption{Applied cuts for the $Z\to \bar\nu\nu$ decay channel at a center of mass energy 
	$\sqrt{s}=\SI{13}{\TeV}$.}
	\vspace*{1em}
	\begin{tabular}{l|c}
		Neutrinos & $p_\text{T}^{\bar\nu\nu} > \SI{150}{\GeV}$ \\
		\multirow{2}{*}{Photon} & $p_\text{T}^\gamma > \SI{140}{\GeV}$, $\abs{\eta^\gamma}<2.37$ \\
		& Frixione isolation $\epsilon_\gamma=0.1, R_0=0.4, n=2$ \\
		Jets & anti-$k_\text{T}$, $D=0.4$, $p_\text{T}^\text{jet} > \SI{30}{\GeV}$, $\abs{\eta^\text{jet}}<4.5$ \\
		Separation & $\Delta \Phi(\gamma,\bar\nu\nu) > 0.6, \Delta \Phi(j,\bar\nu\nu) > 0.5$
	\end{tabular}
\label{tab:cuts-13TeV-Znunu}
\end{table}

\begin{figure}
	\includegraphics[width=\columnwidth]{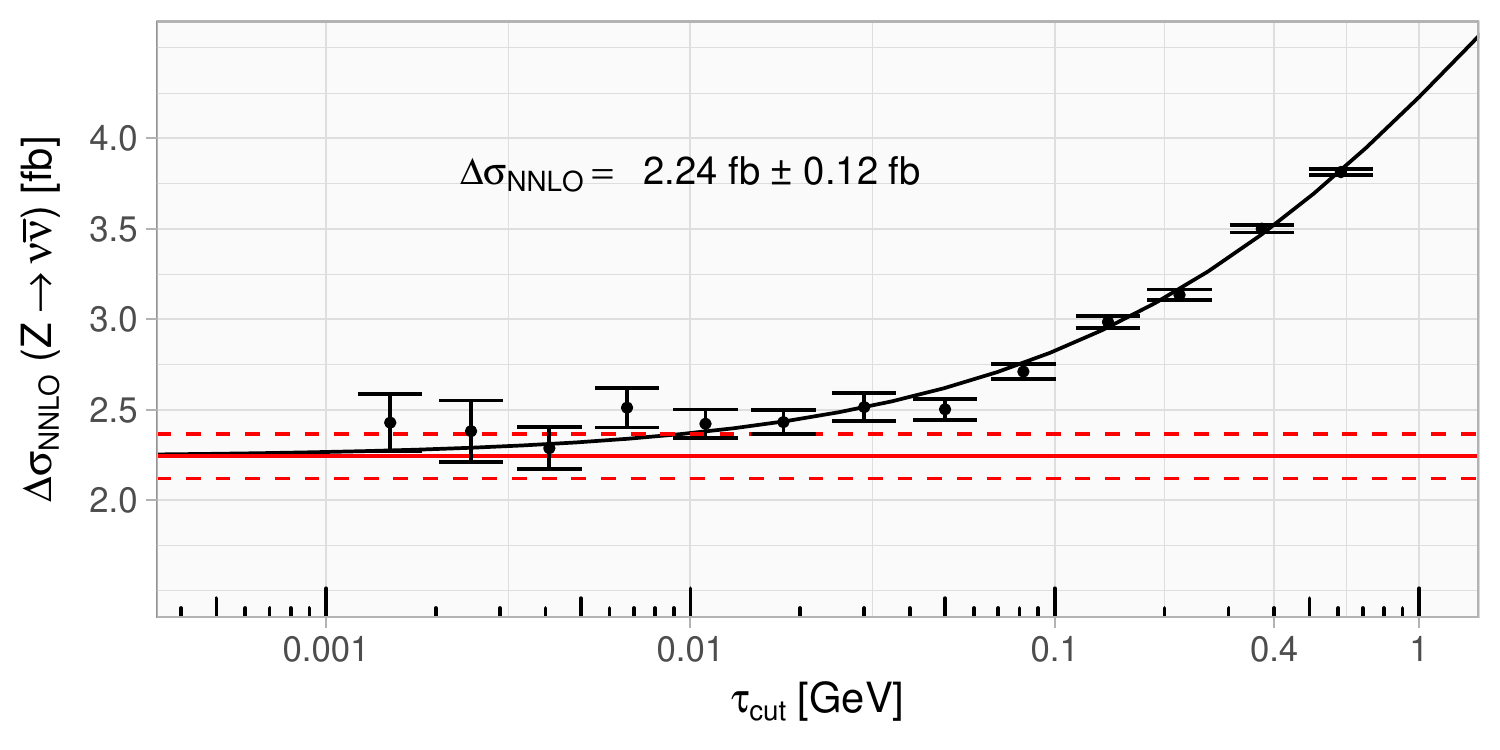}
	\caption{$\taucut$ dependence of the \NNLO{} cross section 
	coefficient in the neutrino decay channel at $\sqrt{s}=\SI{13}{TeV}$ with cuts given in 
	\cref{tab:cuts-13TeV-Znunu}. The solid black line is a 
	fit to the expected asymptotic behavior in \cref{eq:deltannloasymptote}. The solid red line 
	shows the asymptotic value for $\taucut\to0$ of $\Delta\sigma_\text{NNLO} = 
	\SI[separate-uncertainty = true]{2.24\pm0.12}{fb}$. The dashed red lines show the 95\% 
	uncertainty band of about $\SI{0.12}{fb}$.}
	\label{fig:13TeV-Znunu-taucut-nnlocoeff}
\end{figure}

We must first re-establish the appropriate range of $\taucut$ values that represents asymptotic behavior for our 
new set of cuts with $\sqrt{s}=\SI{13}{TeV}$.  As before, we compute the $\tau_\text{cut}$ dependence of the 
\NLO{} cross section coefficient and compare it to the C.-S. obtained result.  We find that the 
same 
range of $\tau_\text{cut}$ leads to asymptotic behavior, and values of 
$\tau_\text{cut}\simeq\SI{0.1}{\GeV}$ already lead to less than a percent difference from the extrapolated 
value. Again, our fitted value of \SI{20.45(1)} is in perfect agreement with the C.-S. result of 
\SI{20.439(3)}{fb}.  We find a similar situation at NNLO, as illustrated in \cref{fig:13TeV-Znunu-taucut-nnlocoeff},
which shows the $\taucut$ dependence of the \NNLO{} coefficient.  Using the 
fitted extrapolated value of \SI[separate-uncertainty]{2.24(12)}{fb} allows us to calculate the 
total 
\NNLO{} cross section as \SI[separate-uncertainty]{86.09(12)}{fb}.

Our results for the cross section under this set of cuts, at each order in QCD, are shown in \cref{tab:13tevnunucross}.
At each order we have also computed the corresponding scale uncertainty, obtained using the 7-point variation
discussed in \cref{par:parameters}.  We see that these scale uncertainties are reduced when going from NLO
to NNLO but remain well above the residual 95\% confidence level uncertainty that results from performing the fit
to the $\tau_\text{cut} \to 0$ limit.  We therefore conclude that the use of the jettiness slicing method, and our
extrapolation procedure, does not lead to any systematic effect of phenomenological relevance.  In \cref{tab:13tevnunucross}
we also indicate the expectation for the jet-vetoed cross section ($N_\text{jet}=0$), although this 
prediction is susceptible to large logarithmic contributions that may require 
resummation, see for example a recent treatment of such issues for $W^+W^-$ production~\cite{Dawson:2016ysj}.

\begin{table}[]
	\centering
	\caption{\SI{13}{TeV} cross sections for the neutrino decay channel with cuts 
	given in \cref{tab:cuts-13TeV-Znunu}. For the \NNLO{} result the additional numerical 
	uncertainty corresponds to the $95\%$ confidence level uncertainty reported by the non-linear 
	fit to the asymptotics in \cref{eq:deltannloasymptote}. It implicitly includes the statistical 
	Monte Carlo integration certainty. Scale uncertainties are obtained by 7-point variation 
    rounded to a full percent.}
	\begin{tabular}{l|c|c|c}
		 & $\sigma_\text{LO}$~[pb] & $\sigma_\text{NLO}$~[pb] & $\sigma_\text{NNLO}$~[pb] \\
		 & $\pm~\text{scale var.}$ & $\pm~\text{scale var.}$ & 
		 $\pm~\text{scale var.}\pm~\text{num.}$ \\
		\hline
		$N_\text{jet}\geq0$ & \multirow{2}{*}{ $53.4 \pm 1\%$ } & $82.0 \pm 3\%$ & $86.09 \pm 1\% 
		\pm 0.14\%$ \\
		$N_\text{jet}=0$ &  & $48.8 \pm 4\%$ & $46.42 \pm 2\% \pm 0.26\%$
	\end{tabular}
	\label{tab:13tevnunucross}
\end{table}

\begin{figure}
	\includegraphics[width=\columnwidth]{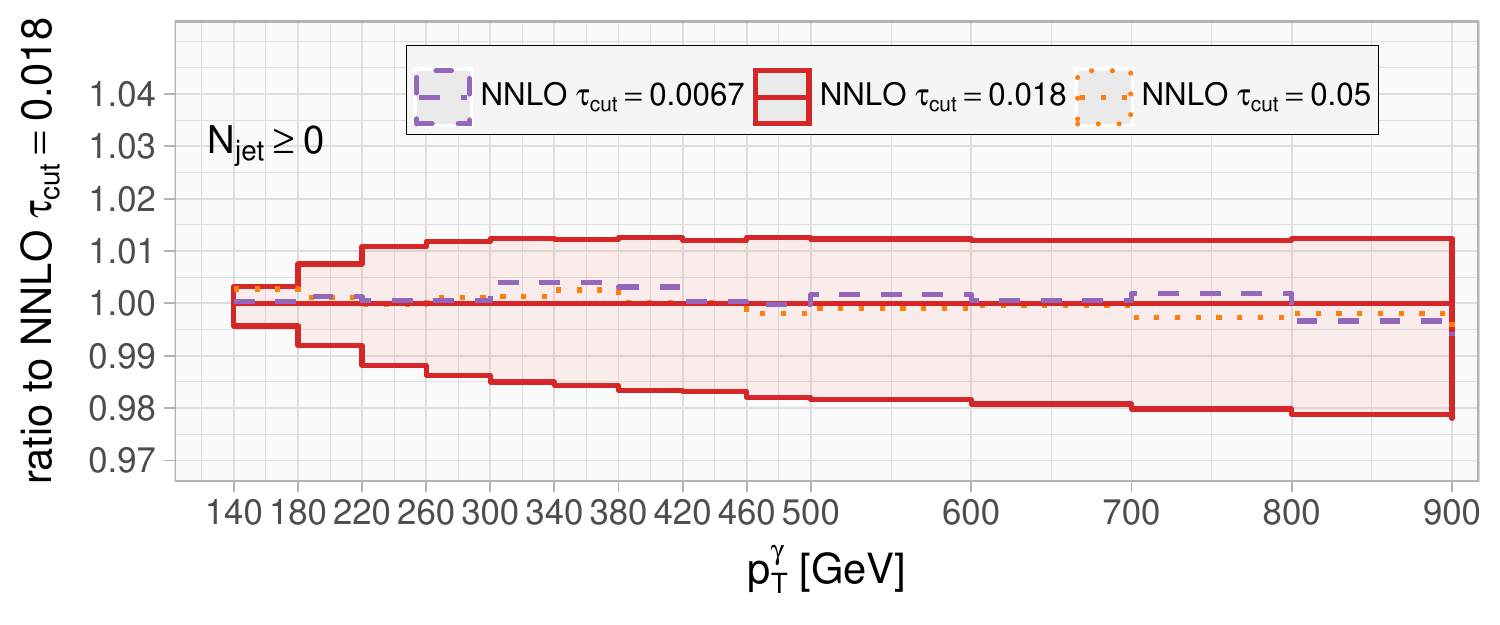}
	\caption{Normalized \SI{13}{TeV} $p_T^\gamma$ \NNLO{} distribution for the neutrino decay 
	channel 
	with different $\tau_\text{cut}$ values in GeV and cuts given in 
		in \cref{tab:cuts-13TeV-Znunu}. The numerical integration uncertainty for the \NNLO{} cross 
		section	results are $\lesssim0.2\%$.	Note that 
		in this \NNLO{} ratio plot the relative uncertainties have to be added. }
	\label{fig:13TeV-Znunu-ptgamma-nnlo-tech}
\end{figure}

We now turn to a study of NNLO predictions for kinematical distributions.   We begin by considering
a differential evaluation of the statistical and systematic numerical uncertainties given by 
integration precision and finite $\tau_\text{cut}$, just as we have already done for the total
cross section.  This is illustrated in
\cref{fig:13TeV-Znunu-ptgamma-nnlo-tech}, where we plot the ratio of the \NNLO{} $p_T^\gamma$ 
distribution with $\tau_\text{cut}$ values of \SIlist{0.05;0.018;0.0067}{\GeV} to the central value of 
the $\tau_\text{cut}=\SI{0.018}{\GeV}$ distribution. For reference we present the scale uncertainty band obtained for 
$\tau_\text{cut}=\SI{0.018}{\GeV}$. 
The first observation 
from the plot is that the residual systematic uncertainty induced by a finite 
$\tau_\text{cut}$ is small compared to the remaining integration uncertainty and the scale 
variation uncertainty. Additionally, the numerical integration uncertainty is small compared to the 
scale variation uncertainty.
Although the scale uncertainty dominates the total, if it is imperative to obtain results with uncertainties
of $0.1\%$ then  a more costly systematic extrapolation of $\tau_\text{cut}\to0$ can certainly be performed.
This could be performed in a manner similar to the total cross section studies presented in \cref{tab:13tevnunucross} and 
\cref{fig:13TeV-Znunu-taucut-nnlocoeff}.  For simplicity, since the systematic residual $\tau_\text{cut}$
dependence is small compared to the scale uncertainty, henceforth all our studies of distributions will
be performed using $\tau_\text{cut}=\SI{0.018}{\GeV}$.

\begin{figure}
	\includegraphics[width=\columnwidth]{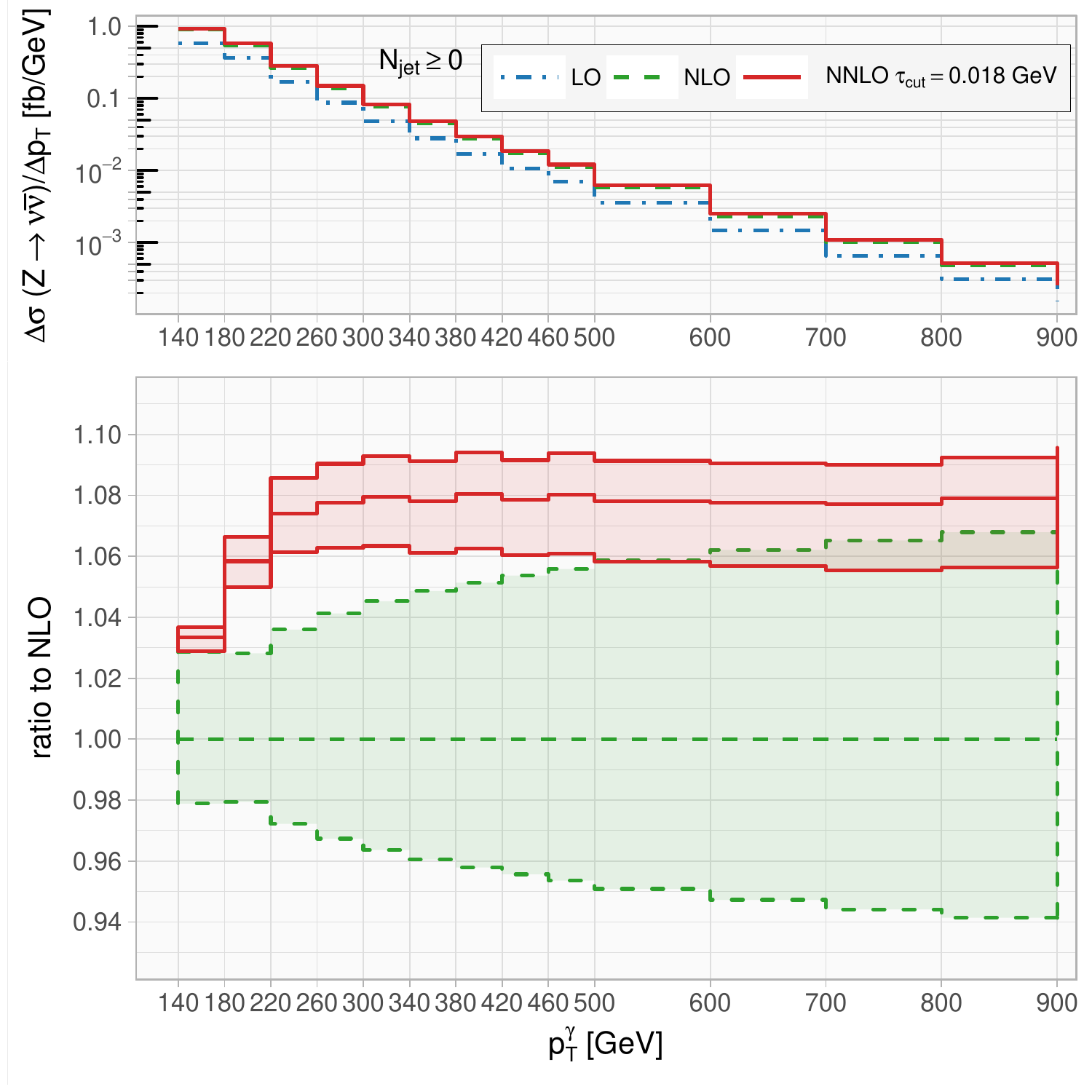}
	\caption{\SI{13}{TeV} $p_T^\gamma$ distribution for the neutrino decay channel with the cuts 
		in \cref{tab:cuts-13TeV-Znunu}. The upper panel shows the differential jet inclusive cross 
		section, while the 
		lower panel shows the ratio to the differential \NLO{} distribution including a scale 
		variation 
		uncertainty band. The numerical integration uncertainty is $\lesssim0.2\%$ for 
		the \NNLO{} result.}
	\label{fig:13TeV-Znunu-ptgamma-nnlo}
\end{figure}

\begin{figure}
	\includegraphics[width=\columnwidth]{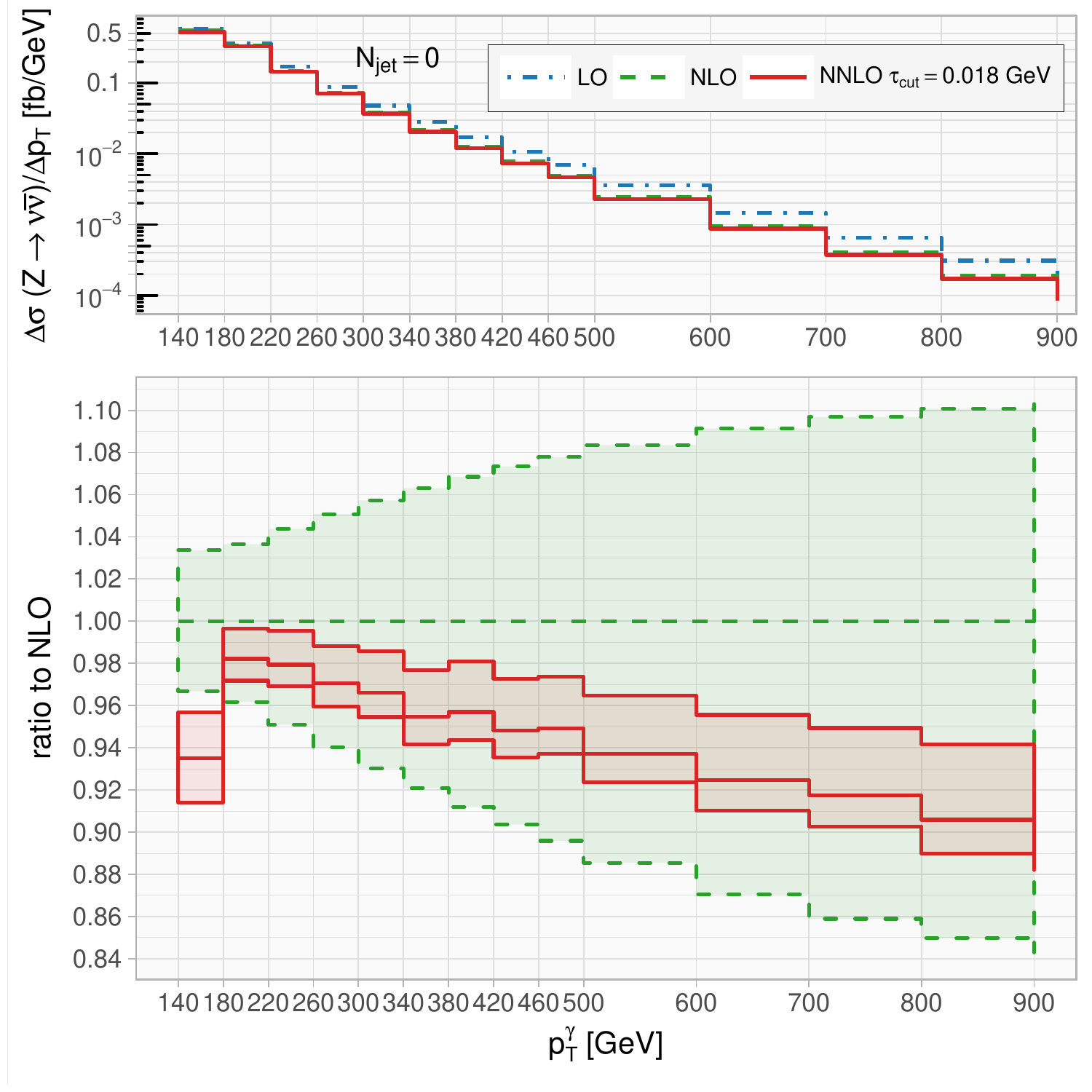}
	\caption{\SI{13}{TeV} $p_T^\gamma$ distribution for the neutrino decay channel with the cuts 
	in \cref{tab:cuts-13TeV-Znunu}. The upper panel shows the differential jet vetoed cross 
	section, while the 
	lower panel shows the ratio to the differential \NLO{} distribution including a scale variation 
	uncertainty band. The numerical integration uncertainty is about \numrange{0.2}{0.5}\%.
	}
\label{fig:13TeV-Znunu-ptgamma-nnlo-jetveto}
\end{figure}

\begin{figure}
	\includegraphics[width=\columnwidth]{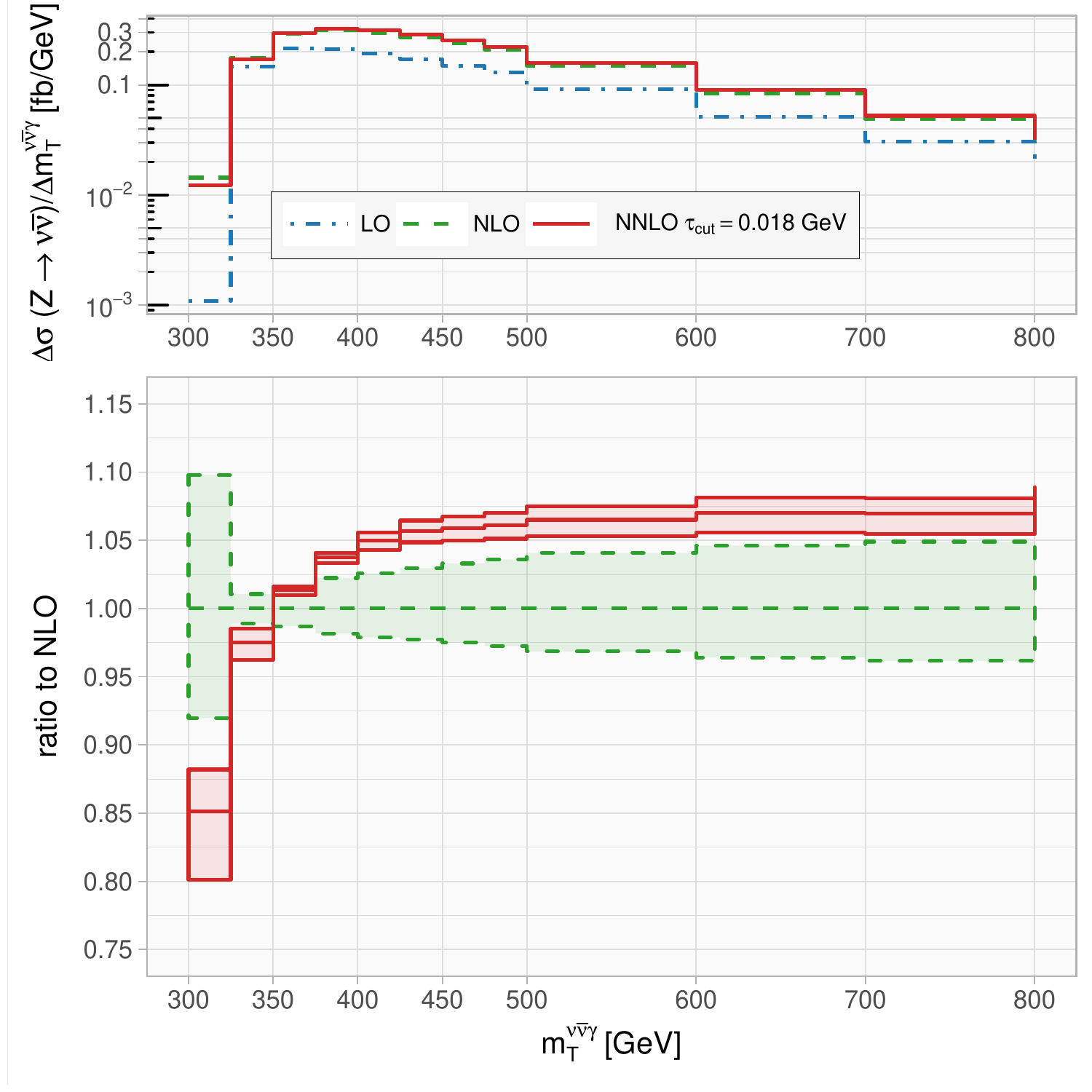}
	\caption{\SI{13}{TeV} $m_T^{\nu\bar\nu\gamma}$ distribution for the neutrino decay channel
	decay with cuts given in \cref{tab:cuts-13TeV-Znunu}. The upper panel shows the differential 
	cross section, while the lower panel shows the ratio to the differential \NLO{} distribution 
	including a scale variation uncertainty band. The numerical integration uncertainty 
	is  $\lesssim0.2\%$ for the \NNLO{} results.}
	\label{fig:13TeV-Znunu-mnunugamma-nnlo}
\end{figure}

In \cref{fig:13TeV-Znunu-ptgamma-nnlo,fig:13TeV-Znunu-ptgamma-nnlo-jetveto,fig:13TeV-Znunu-mnunugamma-nnlo}
we show results for the  jet inclusive and 0-jet exclusive photon transverse momentum distributions, as well as the
$\nu\bar\nu\gamma$ transverse mass distribution, respectively.
\Cref{fig:13TeV-Znunu-ptgamma-nnlo} shows the jet inclusive photon transverse momentum distribution 
at different perturbative orders as well as the $k$-factor relative to the \NLO{} result. As anticipated in the
discussion above, the  numerical uncertainty of the \NNLO{} results of about $0.2\%$ is subleading 
with respect to the 
scale uncertainty band. The impact of the perturbative 
corrections for the new set of \SI{13}{\TeV} cuts is much smaller than for the \SI{7}{\TeV} 
\ATLAS{} analysis. Whereas the corrections for the \SI{7}{\TeV} cuts are 
between $10$ and $20\%$ for $p_T^\gamma<\SI{250}{\GeV}$ \cite{Grazzini:2015nwa}, the perturbative 
corrections for our set of cuts are between $4-8\%$ and are flat for large $p_T^\gamma$.

In \cref{fig:13TeV-Znunu-ptgamma-nnlo-jetveto} we consider the photon transverse momentum 
distribution with a jet veto, where the \NNLO{} corrections increase the cross section by up 
to about $10\%$ at 
$p_T^\gamma\sim\SI{1}{\TeV}$. Finally, we show the transverse mass distribution of the colorless final 
state particles in \cref{fig:13TeV-Znunu-mnunugamma-nnlo}. Again the \NNLO{} corrections are flat 
and about $5$--$10\%$ when sufficiently far away from the production threshold. For these distributions
we have also checked again that the systematic error due to a finite $\tau_\text{cut}$ is small
compared to the other uncertainties.

\subsection{Decay to charged leptons as background to a Higgs signal} \label{sec:Hbackground}

A particularly interesting aspect of the decay channel to electrons (charged leptons) is its
importance as a background to Higgs production with subsequent decay to $Z\gamma$. In
this subsection we consider this case using the cuts given in \cref{tab:cuts-13TeV-Zepem-Higgs}, which
are motivated by the \SI{8}{TeV} 
\ATLAS{} $H\to Z\gamma$ search in ref.~\cite{Aad:2014fia}. Our cuts resemble those for the search 
in the $e^+e^-$ decay channel. For the muon decay channel the cuts in ref.~\cite{Aad:2014fia} are 
altered.
\begin{table}[]
	\centering
	\caption{Applied cuts for $Z\to e^+e^-$ decay channel at a center of mass energy 
		$\sqrt{s}=\SI{13}{\TeV}$ as a background for $H\to Z\gamma$. The cuts are 
		motivated by the \SI{8}{\TeV} analysis in ref.~\cite{Aad:2014fia}.}
	\vspace*{1em}
	\begin{tabular}{l|c}
		Leptons & $p_\text{T}^l > \SI{10}{\GeV}$, $\abs{\eta^l} < 2.47$ \\
		\multirow{2}{*}{Photon} & $p_\text{T}^\gamma > \SI{15}{\GeV}$, $\abs{\eta^\gamma}<1.37$ \\
		& Frixione isolation $\epsilon_\gamma=0.1, R_0=0.4, n=2$ \\
		Jets & anti-$k_\text{T}$, $D=0.4$, $p_\text{T}^\text{jet} > \SI{30}{\GeV}$, 
		$\abs{\eta^\text{jet}}<4.5$ \\
		\multirow{2}{*}{Separation} & $m_{l^+l^-}>m_Z-\SI{10}{\GeV},$ \\
		& $\Delta R(l,\gamma) > 0.3$, $\Delta (l/\gamma,\text{jet}) > 0.3$
	\end{tabular}
	\label{tab:cuts-13TeV-Zepem-Higgs}
\end{table}

\begin{figure}
	\includegraphics[width=\columnwidth]{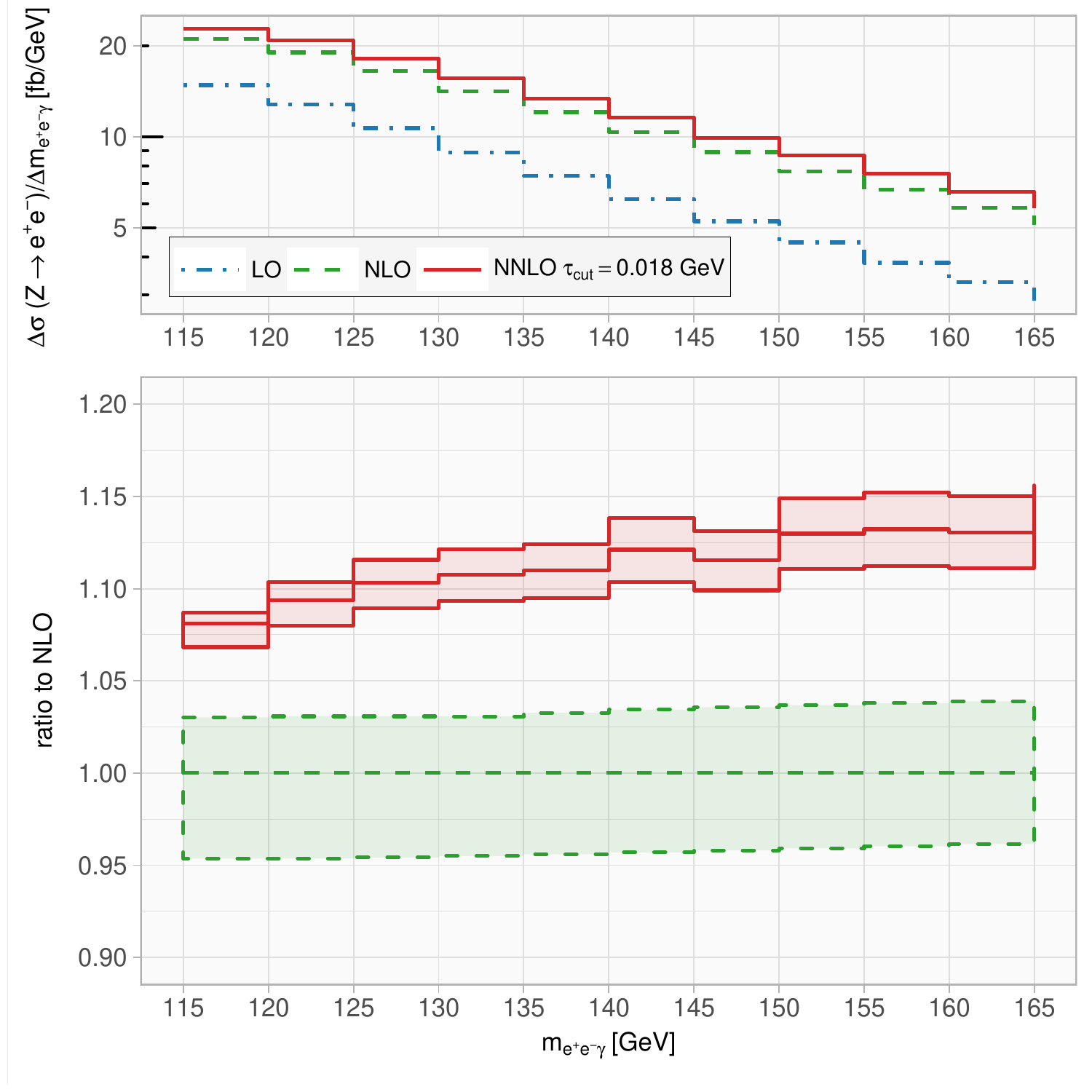}
	\caption{Invariant mass distribution $m_{e^+e^-\gamma}$ of the $e^+e^-\gamma$ system  
		for the electron decay channel with cuts in 
		\cref{tab:cuts-13TeV-Zepem-Higgs}. In the upper panel the absolute distribution is shown, 
		whereas in the lower panel the ratio to the \NLO{} result is displayed. The reported 
		numerical 
		integration uncertainty is about $0.2\%$ for the \NNLO{} results.}
	\label{fig:13TeV-Hbackground-Zepem-mllgamma}
\end{figure}

Since the $H\to Z\gamma$ signal in the charged lepton plus photon invariant mass spectrum  
$m_{l\bar l \gamma}$ will be tiny compared to the 
direct $Z\gamma$ production background an  understanding of the $m_{l\bar l \gamma}$ background 
shape
will be essential.\footnote{We note that there is also an important background contribution from
misidentification of $Z+\text{jet}$ events.} This is despite the fact that experimental analyses 
generally rely on polynomial sideband fits. Nevertheless, perturbative corrections must be 
calculated to make sure that no unexpected shape corrections are introduced until the theory 
uncertainties are under good control at \NNLO{} and beyond. To study this, we show the $m_{l\bar l 
\gamma}$ spectrum in \cref{fig:13TeV-Hbackground-Zepem-mllgamma} and focus on 
the \NNLO{} to \NLO{} $k$-factor: the perturbative \NNLO{} corrections are fortunately relatively 
flat and about $8$--$14\%$ relative to \NLO{}. Since the shape of the spectrum is stable as the perturbative order
increases, it may be beneficial to compare the data (and any resulting polynomial fits) to a theoretical prediction
such as this one. Deviations between the two could thus be used to constrain experimental issues 
such as mis-tagged background events. 

We have also computed the theoretical expectation for the signal process at NNLO under the same set of cuts.
Our predictions for the corresponding $H\to Z\gamma$ signal cross section in gluon fusion production in 
the infinite top-mass limit are given in \cref{tab:13tevepemcross}. As is typical for gluon fusion 
Higgs production, the perturbative corrections are large and a big scale uncertainty remains at 
\NNLO{}. To check whether the signal to background ratio changes noticeably when taking into 
account the \NNLO{} corrections, we also include the background cross section in 
\cref{tab:13tevepemcross}, which has been obtained by adding the bins in the range 
\SIrange{120}{130}{\GeV} from \cref{fig:13TeV-Hbackground-Zepem-mllgamma}. Depending on the 
experimental analysis this range might have to be increased, of course, further suppressing the S/B 
ratio. While the central S/B ratio of about four per-mille increases slightly towards \NNLO{} from 
about $3.5$ per-mille at \NLO{}, the substantial scale uncertainties at \NLO{} already account for such 
a change.
\begin{table}[]
	\centering
	\caption{Inclusive \SI{13}{TeV} cross sections for $H\to Z\gamma$ in gluon fusion in the 
	infinite top-mass limit (signal) and $Z\gamma$ (background) with cuts for the 
	$H\to Z\gamma$ search given in \cref{tab:cuts-13TeV-Zepem-Higgs} in the expected signal region 
	$\SI{120}{\GeV} < m_{l\bar l\gamma} < \SI{130}{\GeV} $.	For the signal cross section our 
	central renormalization and factorization scale is equal to our chosen Higgs mass of 
	\SI{125}{\GeV}.	The \NNLO{} background cross section has an additional numerical uncertainty of 
	$0.2\%$.
	}
	\begin{tabular}{c|c|c|c}
		& $\sigma_\text{LO}$~[fb] & $\sigma_\text{NLO}$~[fb] & $\sigma_\text{NNLO}$~[fb] \\
		& $\pm~\text{scale var.}$ & $\pm~\text{scale var.}$ & $\pm~\text{scale var.}$ \\ 
		\hline & & \\[-1em]
		signal & $0.36^{+26\%}_{-23\%}$  & $0.63^{+20\%}_{-18\%}$ & $0.79^{+10\%}_{-11\%}$ \\[0.5em]
		background & $117.3^{+12\%}_{-15\%}$ & $177.5^{+3\%}_{-5\%}$ & $195.0^{+1\%}_{-1\%}$
	\end{tabular}
	\label{tab:13tevepemcross}
\end{table}

\section{Anomalous couplings and probe for new physics}
\label{sec:anomcoup}

In this section we study anomalous $ZZ\gamma$ and $Z\gamma\gamma$ coupling
contributions introduced by field operators up to dimension 8, requiring
Lorentz invariance and electromagnetic gauge invariance
\cite{Hagiwara:1986vm,Baur:1992cd,Gounaris:1999kf,DeFlorian:2000sg}. The effective 
$Z\gamma Z$ vertex is described by

\begin{multline*}
\Gamma^{\alpha \beta \mu}_{Z \gamma Z}(q_1, q_2, p) = 
\frac{i(p^2-q_1^2)}{\Lambda^2} \Biggl( 
h_1^Z \bigl( q_2^\mu g^{\alpha\beta} - q_2^\alpha g^{\mu \beta}
\bigr) + \\
 \frac{h_2^Z}{\Lambda^2} p^\alpha \Bigl( p\cdot q_2\ g^{\mu\beta} -
q_2^\mu p^\beta \Bigr)
- h_3^Z \varepsilon^{\mu\alpha\beta\nu} q_{2\, \nu} 
- \frac{h_4^Z}{\Lambda^2} \varepsilon^{\mu\beta\nu\sigma} p^\alpha
p_\nu q_{2\, \sigma} \Biggl)\,
\end{multline*}
where $h_i^Z$, $i=1,\ldots,4$ are the effective $Z\gamma Z$ coupling factors and $\Lambda$ is 
conventionally chosen to be the $Z$ boson mass. For a different scale 
the coupling factors have to be scaled accordingly \cite{Baur:1992cd}. The couplings for $i=1,2$ 
are CP-violating, whereas for $i=3,4$ they preserve CP symmetry. The vertex for $Z\gamma\gamma$ can 
be obtained by setting $q_1^2=0$ and replacing $h_i^Z$ by $h_i^\gamma$. Since the 
CP-conserving and CP-violating couplings do not interfere, their sensitivities to anomalous 
couplings are nearly identical and usually only $h_3^V$ and $h_4^V$ are considered in analyses.

The couplings $h_i^{Z,\gamma}$ are zero at tree level in the \SM{} and thus provide a unique
opportunity to probe for new physics. We have implemented the vertices in
our \NNLO{} code to allow for a unified calculation of \NNLO{} $Z\gamma$ production in
the \SM{} and in the presence of anomalous couplings. Whereas previously in experimental analyses
anomalous coupling contributions were calculated at \NLO{} using \MCFMEIGHT{}
and combined with a \NNLO{} \SM{} calculation \cite{Aad:2016sau}, this error prone combination is 
no longer necessary with the additional benefit of having a consistent \NNLO{} prediction.

In general the effect of anomalous couplings is to cause the matrix elements to grow at high 
energies so that it is this region that has the highest sensitivity to their presence. When 
comparing limits one has to be careful though, as limits obtained from different kinematic regions 
can be based on different assumptions on the scale of new physics $\Lambda$ \cite{Green:2016trm}. 
To partially cure this problem and to avoid unitarity violation, a form factor 
that dampens the anomalous coupling contribution at higher energies can be introduced. In our study 
we do not apply any such form factor.

Measurements of $Z\gamma$ production at the \LHC{} that focus on probing anomalous couplings have 
been performed by \CMS{} and \ATLAS{} at $7$ and \SI{8}{TeV}
\cite{Chatrchyan:2011rr,%
	Aad:2012mr,Chatrchyan:2013nda,Aad:2013izg,Chatrchyan:2013fya,%
	Khachatryan:2015kea,Aad:2016sau,Khachatryan:2016yro}.
Previously anomalous  couplings have been constrained by \CDF{} and
\DO{} at the Tevatron \cite{Aaltonen:2011zc,Abazov:2011qp}.
In their latest \SI{8}{TeV} analysis the \ATLAS{} 
collaboration derives limits both using a form factor with a scale of \SI{4}{TeV} and no form 
factor \cite{Aad:2016sau}, while the \CMS{} collaboration uses no form factor 
\cite{Khachatryan:2016yro}. In all cases the obtained limits are of 
the order $10^{-3}$ on $h_3^V$ and $10^{-5}$ on $h_4^V$.  These results set the scale for the 
appropriate ranges of anomalous couplings that we consider shortly.

For the current best limits from \ATLAS{} and \CMS{},
the \ATLAS{} analyses use an exclusive zero-jet selection with cuts of $\pt^\gamma >\SI{250}{\GeV}$ 
for the 
$Z\to \bar ll$ decay, and $\pt^\gamma >\SI{400}{\GeV}$ for the $Z\to \bar\nu\nu$ decay channel. The 
neutrino channel has the highest sensitivity to anomalous couplings due to the larger branching 
fraction. \ATLAS{} and \CMS{} also give measured cross sections, which currently show a huge 
downward fluctuation with respect to the \SM{}. Such a big negative contribution cannot be achieved or explained with the anomalous coupling 
contributions because the couplings need to be small enough to produce a negative interference 
with the \SM{} that is not overwhelmed by the squared anomalous coupling term. Indeed, the 
cross section dependence on the anomalous couplings is almost purely quadratic, which explains the 
high symmetry of the experimental anomalous coupling limits about zero.

\begin{figure}
	\centering
	\includegraphics[width=\columnwidth]{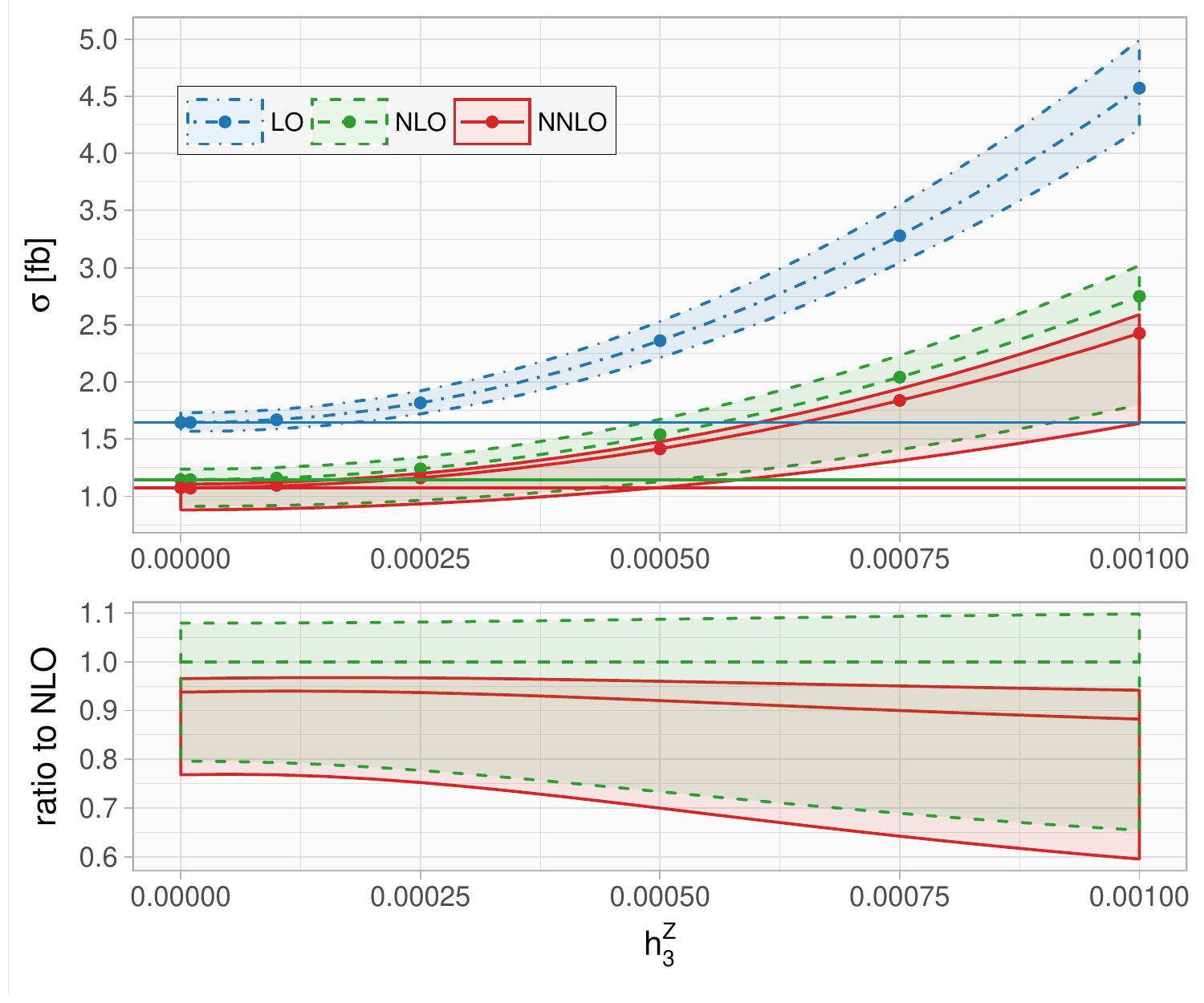}
	\caption{\LO{}, \NLO{} and \NNLO{} cross sections in presence of a non-zero anomalous coupling 
		$h_3^Z$ for the neutrino decay channel with cuts given in \cref{tab:cuts-13TeV-Znunu} and 
		an 
		additional jet-veto as well as $\pt^\gamma>\SI{400}{\GeV}$. The horizontal lines in the 
		upper 
		panel represent the \SM{} cross sections obtained at $h_3^Z=0$.}
	\label{fig:13TeV-Zepem-h3z}
\end{figure}

We do not want to exhaustively cover all anomalous couplings here since their behavior from a 
theoretical viewpoint is mostly similar. We thus give two representative examples and, in each case,
consider ranges of anomalous couplings that directly cover the current limits discussed above.
First we study the dependence of the cross section on just $h_3^Z$ as a one-dimensional example
of the impact of  \NNLO{} corrections on anomalous coupling limits.
\Cref{fig:13TeV-Zepem-h3z} shows the effect of  $h_3^Z$  on the cross section in the neutrino 
decay channel at  \SI{13}{\TeV}. We apply the \SI{13}{\TeV} neutrino channel cuts given in 
\cref{tab:cuts-13TeV-Znunu} with an additional zero-jet veto and a photon transverse 
momentum requirement of $p_T^\gamma>\SI{400}{GeV}$. The change in the central prediction due to the
\NNLO{} corrections is relatively small, around  $5$--$10\%$ over the whole range of 
$h_3^Z$, and negative. However, accounting for the effect of the entire scale uncertainty band allows 
for negative corrections of up to $20$--$40\%$.  Even more interesting is to examine how these
perturbative corrections translate into a change in the estimated limit on $h_3^Z$.

\begin{figure}
	\centering
	\includegraphics[width=\columnwidth]{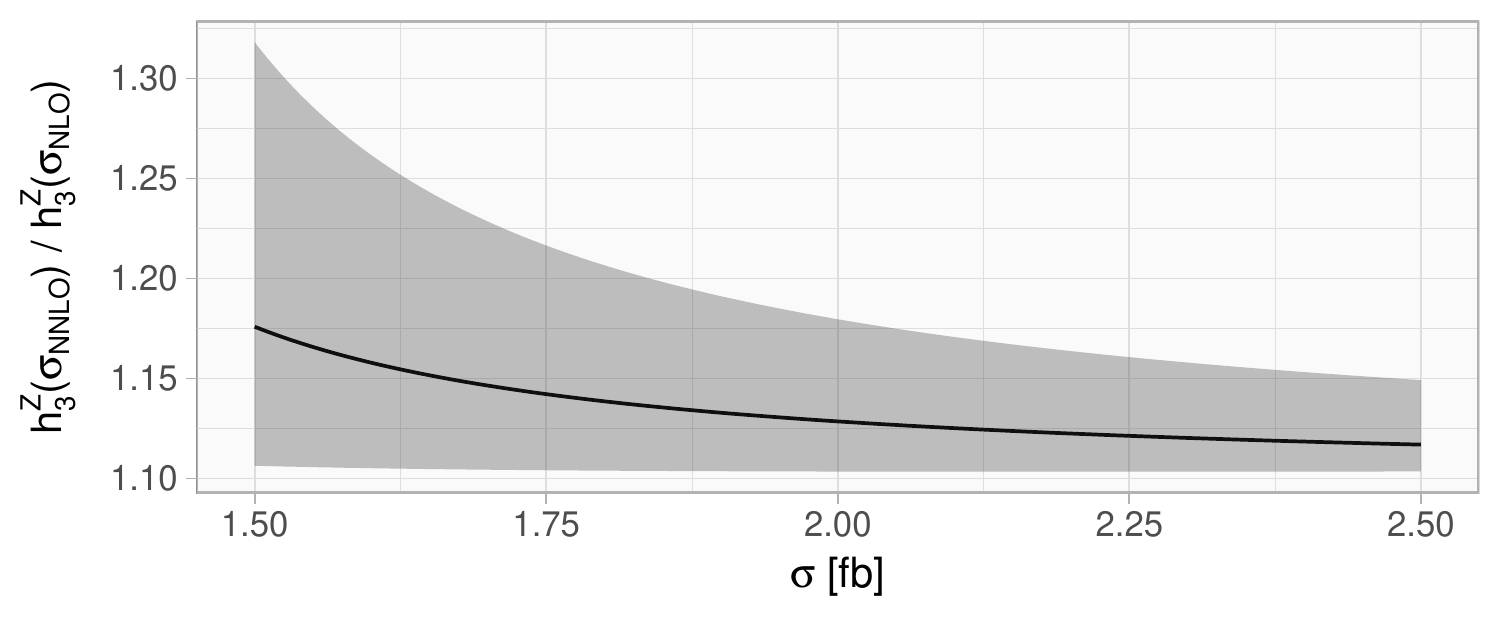}
	\caption{The change in the limit on the anomalous coupling $h_3^Z$ when going from \NLO{} to \NNLO{} as explained in the 
	text. The grey band is obtained by using the scale uncertainty in the theoretical prediction. }
	\label{fig:13TeV-Zepem-h3z-change}
\end{figure}

To directly quantify the change in the estimated limit after accounting for the \NNLO{} corrections
we use the following procedure. We fit a 
quadratic function to describe the dependence of the cross section $\sigma$ on the anomalous coupling $h$, 
$\sigma(h)=\sigma^\text{SM} + h\sigma^\text{interf.} + h^2\sigma^\text{sq.}$.  Here  
$\sigma^\text{SM}$ is the \SM{} cross section and the terms proportional to $h$ and $h^2$ represent the 
effects of interference with the anomalous coupling term and the anomalous coupling squared term, 
respectively. Inverting the equation gives the value of the anomalous coupling as a function of the cross 
section that is measured. We then plot the ratio of the \NNLO{} prediction of 
$\sigma_\text{NNLO}(h)$ to the \NLO{} prediction $\sigma_\text{NLO}(h)$. This number represents the 
amount the limit on an anomalous coupling is changed by the \NNLO{} 
corrections. For example, for $h_3^Z$ this amounts to a loosening of the limit by $10$--$15\%$
using our the central scale choice, driven by the fact that the \NNLO{} corrections are negative.  
We do the same with the envelope given by the scale variation in order to derive an uncertainty band for the 
effect on the limit. The result of this procedure is displayed in \cref{fig:13TeV-Zepem-h3z-change}. 
Other anomalous couplings show a similar behavior, with the \NNLO{} corrections leading to a similar
impact on both the central value that can be extracted and the associated scale uncertainty.

\begin{figure}
	\centering
	\includegraphics[width=\columnwidth]{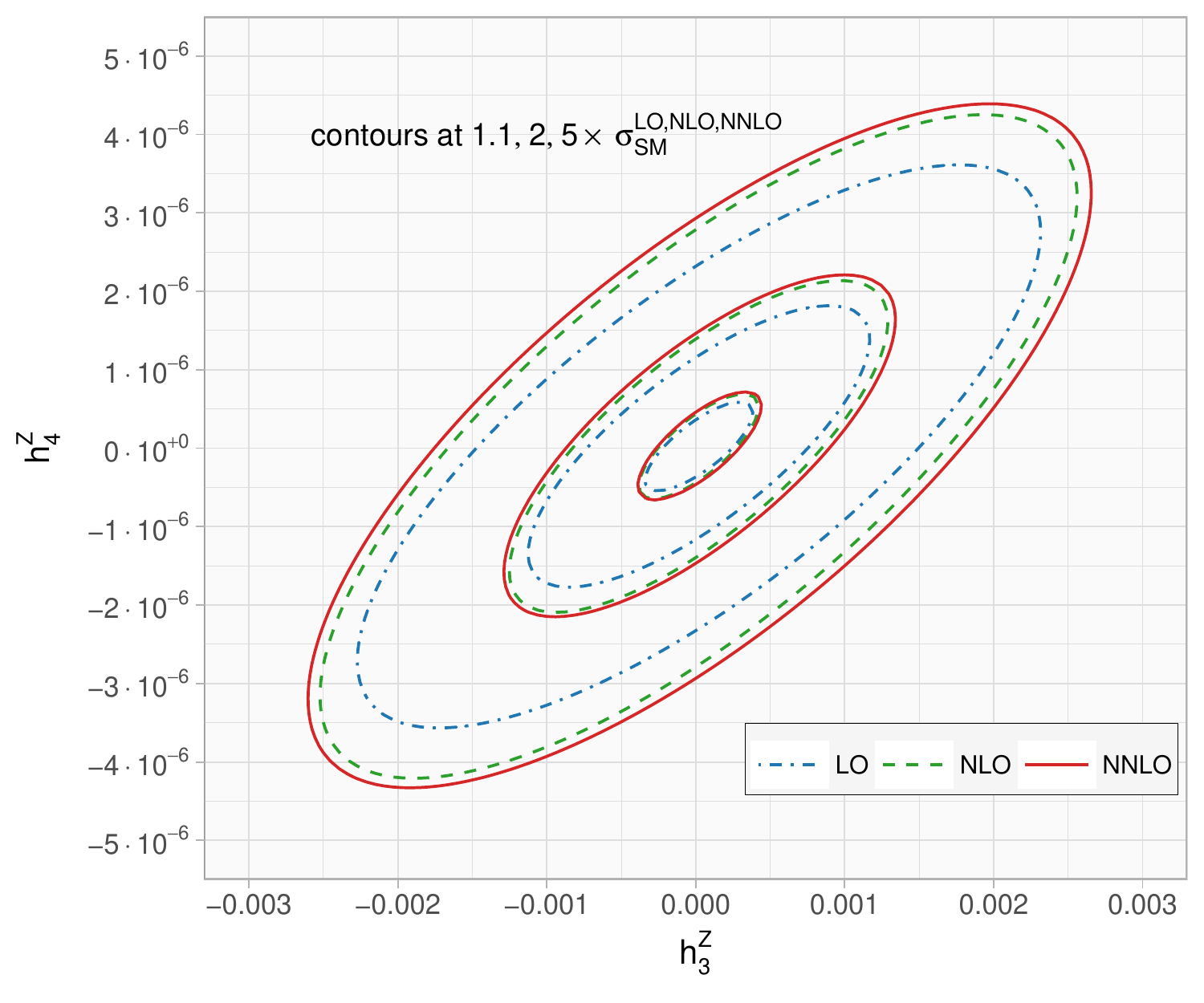}
	\caption{Two dimensional anomalous coupling contour plot in $h_3^Z,h_4^Z$. For each 
	perturbative order the 
	contour for $1.1$, $2$ and $5$ times the corresponding SM prediction is shown. }
	\label{fig:13TeV-Zepem-anomcoup-2D}
\end{figure}

Having outlined the procedure, we now turn to a two-dimensional analysis performed in the 
$h_3^Z,h_4^Z$ plane. In  \cref{fig:13TeV-Zepem-anomcoup-2D} we show a contour 
plot of $\sigma(h_3^Z,h_4^Z)$ at \LO{}, \NLO{} and \NNLO{} with contour lines assuming a cross 
section that is $1.1$, $2$ and $5$ times the \SM{} prediction at the corresponding perturbative 
order. The axis limits are chosen to roughly match the current 2D limits on $h_3^Z,h_4^Z$; see 
figure 14 in ref.~\cite{Aad:2016sau}.
For the plot, we fit the calculated $\sigma(h_3^Z,h_4^Z)$ cross section at interpolation 
points to the functional form $\sigma_\text{SM} + c_1 h_3^Z + c_2 h_4^Z + c_3 h_3^Z h_4^Z + c_4 
h_3^Z h_3^Z + c_5 h_4^Z h_4^Z$. The 2D plot essentially confirms the findings of the 
one-dimensional analysis. The \NNLO{} corrections to the cross section in the presence of
anomalous couplings are relatively small (for the used central scale), as in the \SM{}, but nevertheless
they should be included in order to provide a consistent extraction of limits at \NNLO{}.

\section{Conclusions and outlook}
\label{sec:conc}

We have presented a calculation of $Z\gamma$ production at \NNLO{}, fully including the neutrino and charged 
lepton decays, and in the latter case also including the virtual photon contribution. 
Since we are using the jettiness slicing method, the recent publication of power 
corrections \cite{Moult:2016fqy,Boughezal:2016zws} for Drell-Yan type color singlet processes
motivated a study of their impact on our calculation. For $Z\gamma$ production, in addition to the 
Drell-Yan $s$-channel type of contribution, also a $t$-channel diagram contributes, where the 
photon is not radiated from the final state charged leptons.
Since the existing power corrections are calculated solely for the Drell-Yan type contribution 
their inclusion 
does not dramatically improve the $\tau_\text{cut}$ dependence, unless specific cuts are employed. 
We implemented the boosted definition of the jettiness observable as the 
relevant slicing parameter and found the improvements to be moderate for the cuts used in 
$Z\gamma$ analyses with relatively central $Z\gamma$ systems. Since the improvement is consistent 
and comes essentially for free, we used this definition throughout our study.

For the validation of our ingredients and the implementation itself, we compared with the 
\NNLO{} results in ref.~\cite{Grazzini:2015nwa} and found good agreement in the neutrino decay 
channel but only limited agreement in the charged lepton decay channel. Our matrix elements have 
been 
checked with comparisons against other codes as described in the text. To obtain reliable estimates 
for our cross section predictions we used an extrapolation procedure of 
$\tau_\text{cut}\to0$ determined by the form of the leading jettiness power corrections. We note that if we had used a value 
of $\tau_\text{cut} \sim \SI{0.1}{\GeV}$, we would have found agreement with the existing results 
for both channels, within uncertainties. The difference could be due to the value of the slicing 
parameter in the previous calculation.

For \SI{13}{\TeV} \SM{} phenomenology, we studied $Z\gamma$ in the neutrino decay channel in 
differential distributions with cuts similar to an upcoming \ATLAS{} analysis. We have shown that 
by using a sufficiently low value of $\tau_\text{cut}$, the systematic error made is small compared 
to the numerical integration uncertainty in phenomenological applications, which again is small 
compared to the residual scale uncertainty. For the charged lepton decay channel we envisioned our 
direct $Z\gamma$ production as a background to $H\to Z\gamma$ production. We studied the invariant 
mass distribution of the $Z\gamma$ system and have shown that \NNLO{} effects are relatively flat 
at about $8$--$14\%$ between the Higgs mass and \SI{165}{\GeV}.

Finally, we presented results of our implementation of the $Z\gamma\gamma$ and $ZZ\gamma$ anomalous 
couplings at \NNLO{}. We considered a one-dimensional example and a two-dimensional contour plot 
in the region of current experimental limits. The perturbative corrections for the jet-vetoed set 
of cuts with $p_T^\gamma>\SI{400}{\GeV}$ are about $5$--$10\%$ and will allow for a more precise 
extraction of anomalous coupling limits. Our implementation, which will be released publicly in 
\MCFM{}, will allow for a unified \NNLO{} \SM{} and anomalous couplings analysis.

Diboson production, and specifically $Z\gamma$ production, belongs to a class of processes with 
large 
perturbative corrections and severely underestimated scale uncertainties at \NLO{}. This is due to 
the fact that up to \NNLO{} new partonic channels open up. One thus hopes that for the first time 
with the inclusion of \NNLO{} corrections
the scale uncertainty estimate is reliable, even though the new channels at \NNLO{} are only in 
leading order in their partonic sense. Including higher order effects, for example for the $gg$ 
channel as has been done in diphoton production \cite{Campbell:2016yrh}, is possible 
since all 
ingredients are available. A full N$^3$LO calculation could give a definite answer and should 
be on theorists' agenda anyway, since the uncertainties of results from \ATLAS{} and \CMS{} are 
already competing with the estimated \NNLO{} theory precision. 

\acknowledgments
We thank Roberto Mondini for calculating and providing optimized tree amplitudes. We 
also thank Ian Moult and Al Goshaw for useful discussions regarding the power corrections and 
experimental analysis respectively. 
Support provided by the Center for Computational Research at the University
at Buffalo. C.W. is supported by a National Science Foundation CAREER award
PHY-1652066.
This manuscript has been authored by Fermi Research Alliance, LLC under Contract No. DE-AC02-07CH11359 with the U.S.
Department of Energy, Office of Science, Office of High Energy Physics. The U.S. Government retains and the publisher,
by accepting the article for publication, acknowledges that the U.S. Government retains a non-exclusive, paid-up,
irrevocable, world-wide license to publish or reproduce the published form of this manuscript, or allow others to do so,
for U.S. Government purposes.

\bibliographystyle{JHEP}
\bibliography{VV}

\end{document}